\documentclass[twocolumn,showpacs,preprintnumbers,amsmath,amssymb]{revtex4}
\usepackage{graphicx}
\usepackage{color}

\usepackage{CJK}

\usepackage{amsmath}
\numberwithin{equation}{section}

\usepackage{bm}
\usepackage[normalem]{ulem}
\usepackage{color}
 \newcommand{\bew}{\begin{widetext}}
  \newcommand{\ew}{\end{widetext}}
     
          \newcommand{\dm}{\delta m}
          \newcommand{\dg}{\delta g}
\newcommand{\rp}{r_{_\perp}}
\newcommand{\ls}{\ell_s(\rp)}

 \newcommand{\kbt}{{k_BT}}
          \newcommand{\vb}{{\langle v_x(\br,t)\rangle}}
   \newcommand{\cvv}{\langle \delta v_x(\br+\bR,t+T)\delta v_x({\mathbf{R},T})\rangle}
 \newcommand{\nn}{\nonumber}
 \newcommand{\tc}{\tau_{_{\rm corr}}}

\newcommand{\ii}{{\rm i}}
\newcommand{\mup}{\mu_{_\perp}}

\newcommand{\xpe}{\xi_{_\perp}}
\newcommand{\xpa}{\xi_{_\parallel}}
\newcommand{\nupe}{\nu_{_\perp}}
\newcommand{\nupa}{\nu_{_\parallel}}

\newcommand{\bq}{\mathbf{q}}
\newcommand{\bv}{\mathbf{v}}

\newcommand{\bR}{\mathbf{R}}
\newcommand{\bQp}{\mathbf{Q}_{_\perp}}
\newcommand{\br}{\mathbf{r}}
\newcommand{\brp}{\mathbf{r}_{_\perp}}
\newcommand{\bqp}{\mathbf{q}_{_\perp}}
\newcommand{\bff}{\mathbf{f}}

\newcommand{\al}{\alpha}

\newcommand{\sep}{ \ \ \ , \ \ \ }

\newcommand{\beq}{\begin{equation}}
\newcommand{\eeq}{\end{equation}}
\newcommand{\beqn}{\begin{eqnarray}}
\newcommand{\eeqn}{\end{eqnarray}}
\newcommand{\pp}{\partial}
\newcommand{\dd}{{\rm d}}
\newcommand{\ee}{{\rm e}}

\newcommand{\cO}{{\cal O}}

\newcommand{\cP}{{\cal P}}

\newcommand{\vnab}{\mathbf{\nabla}}

\def\rf#1{(\ref{#1})}

\begin{document}
%\begin{CJK*}{CNS1}}{
% number in the upper righthadn corner of the title page in preprint mode.
%\preprint{}

\title{The Order-disorder Transition in Incompressible Polar Active Fluids with an Easy Axis}
\author{Leiming Chen}
	\email{leiming@cumt.edu.cn}
	\affiliation{School of Materials Science and Physics, China University of Mining and Technology, Xuzhou Jiangsu, 221116, P. R. China}
	\author{Chiu Fan Lee}
	\email{c.lee@imperial.ac.uk}
	\affiliation{Department of Bioengineering, Imperial College London, South Kensington Campus, London SW7 2AZ, U.K.}
	\author{John Toner}
	\email{jjt@uoregon.edu}
	\affiliation{Department of Physics and Institute  for Fundamental Science, University of Oregon, Eugene, OR $97403$}

	\begin{abstract}
Dry active matter in an anisotropic medium is of experimental relevance, and the interplay between anisotropy and the dynamics of the active matter remains under-explored. Here, we derive the hydrodynamic equations of a generic dry polar active fluid that preferentially flows along a particular axis induced by the anisotropy of the medium. We then study its critical behavior at the order-disorder transition in which the symmetry between ``forward" and ``back" along the special axis is spontaneously broken. We obtain the critical static and dynamic exponents, mean velocity, and two point correlation functions  exactly in three dimensions, and to two-loop level  in two dimensions,  by mapping our class of systems to the  equilibrium Ising model with dipolar interactions.
	\end{abstract}
	
\maketitle

\section{Introduction}
The study of ``Active matter"  -  systems consisting of self-propelled agents - and   the collective properties  of such systems,  has been at the forefront of non-equilibrium physics and biological physics for the past two decades \cite{Vicsek1995,toner_annphys05, marchetti_rmp13}.  This research has  revealed that Active matter can be fundamentally different from equilibrium systems, as exemplified by the emergence of a new state of matter that corresponds to the moving phase of  generic  dry polar active fluids in an isotropic medium \cite{Vicsek1995, toner_prl95, toner_pre98}. However, dry active fluids in an {\it anisotropic} environment are also of experimental interest. For instance, anisotropy can be introduced by  patterning the substrate that a two-dimensional (2D) active fluid, such as a collection of self-propelled particles \cite{Bricard2013}, moves on; or by stretching a gel that a three-dimensional (3D) active fluid, such as motile cells \cite{zaman_pnas06}, move within. 

In this paper, we focus on active fluids in which the active particles move preferentially parallel 
 to a certain direction, which we refer as the ``easy axis'', and will denote throughout this paper as $x$. We will denote directions orthogonal to $x$ as $\brp$ in $d=3$, and as $y$ in $d=2$. When driven by an external field aligned  with the easy axis, this active  system has been shown to possess rich phase behavior \cite{katz_prb83,janssen_zpb86,leung_jstatphys86}.

 In this paper, we consider systems {\it without} such an external field; that is, one  in which  the underlying dynamics has ``up-down" symmetry along the easy axis. That is,  if the easy axis is, e.g., vertical, the active
particles are equally likely {\it a priori} to move either up  or down along this axis.
We show here that such a system  can break this up-down symmetry {\it spontaneously}  at a non-equilibrium phase transition.    We find  that, in the incompressible limit,   this transition is continuous, and the corresponding critical behavior belongs to the universality class of the  equilibrium, purely relaxational (i.e., ``TDGL") Ising model with dipolar interactions \cite{aharony_ising73, brezin_prb76, folk_zpb77}. Through this mapping we are able to obtain the  exact scaling behavior of this transition in spatial dimension $d=3$, and  estimates of the  critical exponents in $d=2$, simply by using the results of \cite{aharony_ising73, brezin_prb76, folk_zpb77}.

In this mapping, the local component  $v_x(\br,t)$ of the velocity of the  self-propelled particles along the easy axis plays the role of the local magnetization $M(\br,t)$ of the dipolar Ising model.
We here predict the critical behavior of the mean velocity $\vb$, which is the order parameter of our transition, and perfectly analogous to the mean magnetization $\langle M\rangle$  of the dipolar Ising ferromagnet, near the order-disorder transition. We also use this connection to dipolar systems to show that, in both three and two dimensions,
 the system  is characterized by {\it two} distinct correlation lengths $\xi_{_{\perp, \parallel}}$ (where here and throughout this paper we use $\parallel$ ($\perp$)  to denote directions along (perpendicular to) the easy axis),  and a correlation time $\tau_{_{\rm corr}}$, all three of which diverge in a universal way as the transition is approached.

These correlation lengths can most easily be extracted from two point correlations of  the fluctuation $\delta v_x(\br,t)$ of the local speed $v_x(\br, t)$   about its mean value:

\beq
\delta v_x(\br,t)\equiv v_x(\br, t)-\langle v_x\rangle \,.
\label{del v def}
\eeq
Obviously, on the {\it disordered} side of the transition, where $\langle v_x\rangle=0$, $\delta v_x$ and $v_x$ are identical.

 The  two point correlation function
 \beq
 C(\br,t)\equiv\cvv \,,
 \label{cvvdef}
 \eeq
 in the system,  with the spatial separation $\br\equiv(x, \brp)$,  takes on different forms in the
 ``critical" regime, in which
 all three of the conditions
 \beq
 x\ll\xpa \sep r_{_\perp}\ll\xpe \sep t\ll\tc
 \label{crit reg}
 \eeq
are satisfied, and in the ``non-critical" regime, in  which one or more of the conditions \rf{crit reg} is violated. We obtain exact results for $C(\br,t)$ in the non-critical regime in both  $d=2$ and  $d=3$, and exact results in the critical regime in $d=3$. In $d=2$, we obtain a scaling form for $C(\br,t)$ in the critical regime.

We'll now summarize these results, starting with the $d=3$ case, and then turning to $d=2$.

\subsection{Results in $d=3$}

In $d=3$, the mean velocity of the active agents (which is the perfect analog of the magnetization in a dipolar ferromagnet) in the ordered phase near the transition obeys the {\it exact, universal} scaling law: 
 \beq
 \langle v_x \rangle\propto |p-p_c|^{1/2} \bigg|\ln\left(\Big|{p-p_c\over p_c}\Big|\right)\bigg|^{1/3} \ , \ \  (d=3) \,,
 \label{OPd=3}
 \eeq
where $p$ is the experimental control parameter (which could be, e.g., the density of self-propelled agents, the concentration of ATP in a biological system, chemical concentration in a system of Janus particles, etc.) which is tuned to go through the transition, and $p_c$ is the value of that parameter at which the transition occurs.

The correlation lengths $\xpe$ and $\xpa$, and the relaxation time $\tc$,
diverge as the transition is approached  (i.e., as $p\to p_c$ {\it from either side}) according to the  {\it exact, universal} scaling laws:
 \beq
\xpe \propto |p-p_c|^{-1/2} \bigg|\ln\left(\Big|{p-p_c\over p_c}\Big|\right)\bigg|^{1/6} \ , \ \ (d=3)
\,,
\label{xilog_perp}
\eeq
and
 \beq
\xpa \propto  \tc\propto |p-p_c|^{-1} \bigg|\ln\left(\Big|{p-p_c\over p_c}\Big|\right)\bigg|^{1/3}  \ , \ \ (d=3)
\,.
\label{xilog_para}
\eeq

In the non-critical regime, in which either $|\brp|\gg\xpe$ or $|x|\gg\xpa$, or both, the equal-time correlation function is given by
\begin{eqnarray}
C(\br,0)={f_{_{3D}}(\theta)\over r^3}
\label{eq:vx3I}
\end{eqnarray}
where $\theta=\tan^{-1}\left({r_{\perp}\over x}\right)$ is the polar angle of $\br$ from the $x$-axis, and, near the transition,
\beqn
f_{_{3D}}(\theta)=
{D_x\xpe\left(2\alpha\cos^2\theta-\sin^2\theta\right)\over 4\pi\sqrt{\mup w}(\sin^2\theta+\alpha\cos^2\theta)^{5/2}}\,,
\label{f3dI}
\eeqn

Note that this result applies both {\it above} the transition (i.e., for $p>p_c$) and below it (i.e., for $p<p_c$).

In \rf{f3d}, $\xpe$ is the perpendicular correlation length introduced earlier, and $D_x$,
$\mu_\perp$, $w$, and $\alpha$ are phenomenological parameters of our model. The first three of these are non-singular through the transition, while $\alpha$ vanishes like $\xpe^{-2}$. This implies that the region of $\br$ over which the correlation function  $C(\br,0)$ is positive becomes very narrow as the transition is approached. Specifically, positive correlations only occur in the narrow cone $\theta<\sqrt{2\alpha}\propto\xpe^{-1}$, which becomes infinitesimally narrow as the transition is approached.

The equal position-{\it un}equal-time correlation function  is given by 
\beqn
C_{ep}(t)
\approx
\left\{
\begin{array}{ll}
\left({D_x\over16\pi\sqrt{w\mu_\perp^3}}\right)\left[{1\over |t|}\right]\,, &|t|\ll\tc\,,\\
\left({D_x\tc B_t\over16\pi\sqrt{w\mu_\perp^3}}\right)\left[{\exp(-|t|/\tc)\over t^2} \right]\,, &|t|\gg\tc \,,
\end{array}
\right.
\label{Cet3I}
\eeqn
where  $C_{ep}(t)\equiv C({\mathbf 0},t)$, and $B_t$ is a non-universal, $O(1)$ constant.

 In the critical regime, in which all three of the conditions (\ref{crit reg}) are satisfied, the equal-time correlation function is
 \beqn
C({\bf r}, 0)&=&{D_x\over8\pi\mu_{_\perp}|x|}\exp\left[-{r_{_\perp}^2\over4x_0|x|}\right]
  \ , \ \ (d=3)\,,~~~~~~~
\label{rscorrd=3_I}
\eeqn
where we've defined
\beq
x_0\equiv\sqrt{\mu_{_\perp}\over w} \,.
\label{x0def}
\eeq

\vspace{.2in}

\subsection{Results in $d=2$}

In two dimensions, the mean velocity in the ordered phase is given by
\beq
 \langle v_x \rangle\propto |p-p_c|^{\beta}  \,,
 \label{OPd<3}
 \eeq
where the universal exponent $\beta$ can be obtained from an expansion in powers of $\epsilon\equiv3-d$:
 \beq
 \beta=\frac{1}{2}- \frac{\epsilon}{6} +  \left[{1\over4}  \ln \left(\frac{4}{3}\right) +{155\over972} \right]  \epsilon^2 +
O( \epsilon^3) \,.
\label{betaintro}
\eeq

Likewise, the correlation lengths $\xpe$ and $\xpa$ also diverge algebraically:
\begin{eqnarray}
&&\xpe \propto |p-p_c|^{-\nupe} \sep \xpa \propto  |p-p_c|^{-\nupa}
\, ,
\label{xi}
\end{eqnarray}
with the universal exponents $\nupe$ and $\nupa$
 given by
 \beqn
\label{eq:nupe}
\nupe &=& \frac{1}{2}+ \frac{\epsilon}{12} + {1\over4} \left( \ln \left(\frac{4}{3}\right) +{43\over54} \right) \epsilon^2 +
O( \epsilon^3)\\
\label{eq:nupa}
\nupa &=& \zeta\nupe=1+ \frac{\epsilon}{6} +  {1\over2} \left( \ln \left(\frac{4}{3}\right) +{383\over486} \right) \epsilon^2 +
O( \epsilon^3)  \,.  \nn\\
\eeqn

In the first equality of  \rf{eq:nupa}, we have introduced the ``anisotropy exponent"
\beq
\zeta=2-{\eta\over2}=2-{2\over243}\epsilon^2+O(\epsilon^3) \,,
\label{zetaintro}
\eeq
 which plays a crucial role in many other expressions in addition to this one, as we'll see below, as does $\eta$, which has the $\epsilon$-expansion
 \beq
 \eta={4\over243}\epsilon^2+O(\epsilon^3) \,.
 \label{eta}
 \eeq

  The relaxation time $\tau_{_{\rm corr}}$ also diverges algebraically as the transition is approached:
 \beq
 \tau_{_{\rm corr}}\propto|p-p_c|^{-z\nupe}
 \label{taudiv}
 \eeq
 where
\beq
z\nupe=1+ \frac{\epsilon}{6} +  {1\over2} \left[ \ln \left(\frac{4}{3}\right) +{43\over54}+{4c\over243} \right] \epsilon^2 +
O(\epsilon^3)
\,,
\eeq
with the ``dynamical exponent"
 \beq
 z=2+c\eta+O(\epsilon^3)=2+\left({4c\over243}\right)\epsilon^2+O(\epsilon^3) \,.
 \label{zeps_I}
 \eeq
The constant
\beq
c=.92
\label{cnum}
\eeq
 in these expressions was obtained in \cite{folk_zpb77} by numerically evaluating a  complicated multi-dimensional integral.

Unfortunately, we suspect that some of these  $\epsilon$-expansions do {\it not} actually work very well in $d=2$, where $\epsilon=1$. The reason  is that the $O(\epsilon^2)$ terms in \rf{eq:nupe} and \rf{eq:nupa} are actually bigger than the $O(\epsilon)$ term at  $\epsilon=1$. It is well-known \cite{chaikin} that $\epsilon$-expansions typically give {\it asymptotic} series for the exponents. That means the series do not actually converge. The rule of thumb for such series \cite{Orzag} is  that  the most numerically accurate  results are obtained by truncating such series at the last term that is smaller than the previous term. For the usual $O(n)$ model
$d=4-\epsilon$-expansion in $d=3$, where $\epsilon=1$, this  truncation is  at $O(\epsilon^2)$. But for the dipolar problem, and our non-equilibrium one, we should, by this criterion, truncate our expansions for $\nupe$, $\nupa$, $\beta$ at $O(\epsilon)$,  which means our  error bars will  probably be  comparable to those of the usual $4-\epsilon$-expansion for an $O(n)$ model if truncated at $O(\epsilon)$, which are typically  $\pm.07$ for $\nu_\perp$. This would suggest  $\pm.14$ for $\nupa$, and $\pm.02$ for $\beta$.

Similar observations apply to the expansion for the combination $z\nupa$ that determines the divergence of the relaxation time $\tau_{\rm corr}$ in \rf{taudiv},  and $\beta$. The $O(\epsilon^2)$ corrections to the anisotropy exponent $\zeta$ and the dynamical exponent $z$, on the other hand, are so small that it is probably safe to conclude that the zero-th order in $\epsilon$ results $z\approx\zeta\approx2$ are quite accurate; specifically, they are probably correct to $\pm.01$, the error being roughly the size of the $O(\epsilon^2)$ corrections to those exponents. We also expect that the exponent $\eta$ will be very small, since to leading order in $\epsilon$ it is $\sim .02$.

 Keeping the above caveats in mind, evaluating $\nupe$ and $\nupa$ in $d=2$ using \rf{eq:nupe} and \rf{eq:nupa}  by setting $\epsilon=1$ and truncating according to the aforementioned rule of thumb (which for $\nupe$ and $\nupa$ calls for truncation at first order in $\epsilon$)  gives
\beqn
\nupe&=&.58\pm .1 \sep \nupa=1.17\pm .2 \sep \zeta=2\pm .01 \sep
\nn\\
z&=&2\pm .01 \sep  z\nupe=1.2\pm .2 \sep \beta=.3\pm  .02 \,,\nn\\&& \,\,\,\,\,\,\,\,\,\,\,\,\,\,\, \,\,\,\,\,\,\,\,\,\,\,\,\,\,\, \,\,\,\,\,\,\,\,\,\,\,\,\,\,\, \,\,\,\,\,\,\,\,\,\,\,\,\,\,\, \,\,\,\,\,\,\,\,\,\,\,\,\,\,\, \,\,\,\,\,\,\,\,\,\,\,\,\,\,\,  ({\rm in} \,\, d=2)\,,
\label{nu2d}
\eeqn
 where the error bars are determined by the criteria described above.

In the non-critical regime, in which either $|\brp|\gg\xpe$ or $|x|\gg\xpa$, or both, the equal-time correlation function is given by
 \begin{eqnarray}
C({\bf r}, 0)={f_{_{2D}}(\theta)\over r^2} \sep (d=2)
\label{eq:vx4I}
\end{eqnarray}
where
\beqn
f_{_{2D}}(\theta)\equiv
{
\xpe D_x\over \pi\sqrt{\mu_{_\perp} w}
}
{(\alpha\cos^2\theta-\sin^2\theta)\over(\sin^2\theta+\alpha\cos^2\theta)^2}
\,.
\label{f2dI}
\eeqn
In addition, for $|t|\gg\tau_{\rm corr}$, the equal-position correlation function is given by
\beq
C({\bf 0}, t)=
{D_x\over 4\sqrt{\pi^3\mu_{_\perp}w}\xi_{_\perp}}
\left(|t|\over\tau_{\rm corr}\right)^{-{3\over 2}}
\exp\left(-|t|\over \tau_{\rm corr}\right)\,.
\label{eq:vx5I}
\eeq

In contrast to  $d=3$, here in $d=2$  $\mu_{_\perp}$ is
renormalized by the critical fluctuations, and becomes $\xpe$-dependent: $\mu_{_\perp}\propto \xpe^\eta$.
  In addition, in $d=2$ we find  $\alpha\propto \xpe^{\eta-2}$.

Qualitatively, as in $d=3$, this still implies that the window of positive correlations is again a narrow wedge $\theta<\sqrt{\alpha}\propto\xpe^{{\eta\over2}-1}$, which becomes infinitesimally narrow as the transition is approached.

In the critical regime, in which all of the criteria (\ref{crit reg}) are satisfied, we do not have a closed form analytic expression for the correlation functions. However, we do have its scaling form:
 \begin{eqnarray}
C(\br,t)
&=&r_{_\perp}^{2\chi}F\left({|x|\over\rp^\zeta}, {|t|\over\rp^z}\right)
\nn\\
&\propto&\left\{
\begin{array}{ll}
r_{_\perp}^{2\chi},& {|x|\over a_x}\ll \left({r_{_\perp}\over a_{_\perp}}\right)^{\zeta}, {|t|\over\tau_0}\ll \left({r_{_\perp}\over a_{_\perp}}\right)^z\,,\\
|x|^{2\chi\over\zeta},&{|x|\over a_x}\gg \left({r_{_\perp}\over a_{_\perp}}\right)^{\zeta}, {|x|\over a_x}\gg \left({|t|\over\tau_0}\right)^{\zeta\over z}\,,\\
|t|^{2\chi\over z},& {|t|\over\tau_0}\gg\left({r_{_\perp}\over a_{_\perp}}\right)^z, {|t|\over\tau_0}\gg \left({|x|\over a_x}\right)^{z\over\zeta} \,.
\end{array}
\right.
\label{crit corr}
\end{eqnarray}
In  \rf{crit corr}, $a_{x,_\perp}$ are microscopic lengths which exhibit no critical behavior as the  transition is approached (that is, they remain small and finite right up to the transition), and $\tau_0$ is similarly a non-critical microscopic time.

In \rf{crit corr}, the anisotropy exponent $\zeta$, and dynamical exponent $z$ are those given above by (\ref{zetaintro}) and (\ref{zeps_I}), while the ``roughness exponent" $\chi$ is given by
\beq
\chi=\left({1-d\over2}\right)-{\eta\over4}=-1+{\epsilon\over2}-{\epsilon^2\over243}
 +O(\epsilon^3)\,.
 \label{chi}
 \eeq
 Here and throughout this paper, $\br\equiv (x, \brp)$.

The remainder of this paper is organized as follows:  in section \rf{hydro}, we derive the hydrodynamic equations of motion for a uniaxial incompressible flock. In section \rf{mft}, we present the ``mean-field" theory of such systems (i.e., their behavior in the absence of noise and fluctuations. Section \rf{lin} then treats those fluctuations in the linear theory. In section \rf{id}, we identify the relevant {\it non}-linear terms in the equation of motion. Utilizing this, we show in section \rf{map} that these systems map on to an equilibrium Ising ferromagnet with long-ranged dipolar interactions. Section \rf{RGsec}  reviews the RG analysis of that equilibrium Ising ferromagnet, and its predictions for the correlation and response functions. Section \rf{sum} briefly summarizes our results, and discusses possible directions for future research.

\section{Hydrodynamic equations of motion}{\label{hydro}}
We start with the generic equations of motion (EOM) of dry compressible polar active fluids  easy axis, chosen here to be the $x$-axis. The hydrodynamic variables are the coarse grained velocity field $\bv$ and the number density field $\rho$.
The EOM of $\rho$  follows from particle number conservation:
\beq
\pp_t \rho +\vnab \cdot (\rho \bv) =0
\ .
\label{cont}
\eeq
For the EOM of $\bv$, we assume forward-backward symmetry in $v_x$ (i.e., the EOM remains invariant under the simultaneous mappings $v_x \mapsto -v_x$ and $x \mapsto -x$), and rotational invariance within the dimensions perpendicular to the $x$-axis  (in spatial dimensions $d>2$, or invariance under $v_\perp\to-v_\perp$ in $d=2$. The generic EOM with these symmetries are of the form:

\begin{widetext}
\begin{eqnarray}
\pp_t v_x &=&
-\left[\lambda_1v_x\partial_x+
\lambda'_1 (\bv_{_\perp} \cdot\vnab_{_\perp})\right]v_x
-\left[\lambda_2\partial_xv_x+\lambda'_2(\vnab_{_\perp}\cdot\bv_{_\perp})\right]v_x
-\partial_x g(v_{_\perp}^2,v_x^2,\rho)\nonumber\\
&&-K\partial_x\cP-\left(a+b v_x^2\right) v_x
+\left(\mu_1\partial^2_x+\mu_\perp \nabla_{_\perp}^2\right)v_x
+\partial_x(\mu_2\partial_xv_x+\mu'_2\vnab_{_\perp}\cdot\bv_{_\perp})+ f_x \, ,
\label{V_x1}\\
\pp_t \bv_{_\perp} &=& -\vnab_{_\perp}\cP-c\bv_{_\perp}+ \bff_{_\perp} \, ,\label{V_perp}
\end{eqnarray}
\end{widetext}
where only potentially relevant terms are shown, and $g$ is a scalar function that is analytic in the field variables,  and even in $v_x$ and $v_\perp$ as required by the aforementioned symmetries.  Note that the damping term $-c\bv_{\perp}$ reflects  the preference of the active particles for moving parallel to the $x$-axis.  The dimensionless parameter $K\ne1$ in general, since our system is anisotropic, and therefore responds  differently to pressure gradients
along the easy axis than to those perpendicular to it.
The random force terms $\bff$ have the following statistics:
\begin{eqnarray}
&&\langle f_i(\br, t)f_j(\br',t')\rangle\nonumber\\
&=&2\left(D_x\delta_{ij}^x+D_{_\perp}\delta_{ij}^{_\perp}\right)
\delta^d(\br-\br')\delta(t-t')\,,\label{Noise_corr}
\end{eqnarray}
where $\delta_{ij}^x = 1$ if $i=j=x$ and is zero otherwise, and $\delta_{ij}^{_\perp} = 1$ if $i=j\neq x$ and is zero otherwise.

Without fluctuations, the order-disorder transition happens at $a=0$, with an Ising-like spontaneous
symmetry breaking. The order parameter of this transition is $v_x$. For $a>0$
the system is in the disordered state with mean velocity $\langle\bv\rangle=\mathbf 0$, while
for $a<0$ ,the system is in the ordered state with non-zero mean velocity
$\langle v_x\rangle\neq 0, \langle\bv_{_\perp}\rangle=\mathbf 0$.

We now take the incompressible limit by making the pressure term $\cP$ extremely sensitive to density variations. This forces the density to be constant, which, as in incompressible simple fluids,  reduces the continuity equation \rf{cont} to the constraint 
\beq
\vnab \cdot \bv =0 \,.%\\\nonumber\\
\label{divless}
\eeq
\vspace{.1in}

In this limit, the pressure term $\cP$ corresponds to the Lagrange multiplier in the EOM that enforces this incompressibility condition \rf{divless}.  We can calculate the pressure as follows. We first apply
$\partial_x$ and $(\vnab_{_\perp}\cdot)$ on both sides of the equalities in (\ref{V_x1})
and (\ref{V_perp}), respectively;
 we then add the two resultant equations together.

This gives:
\bew
\beqn
\pp_t\vnab \cdot \bv =&-&(K\pp_x^2+\nabla_\perp^2)\cP-a\pp_xv_x-c\vnab_\perp \cdot \bv_\perp-b\pp_x(v_x^3)-\lambda'_1\pp_x[(\bv_\perp\cdot\vnab)v_x]-\lambda'_2\pp_x[(\vnab_\perp\cdot\bv_\perp)v_x]
\nn\\
&-&(\lambda_1+\lambda_2)v_x\pp_x v_x+(\mu_1\pp_x^2+\mu_\perp\vnab^2_\perp)\pp_x v_x+\mu_2\pp_x^3 v_x+\mu_2'\pp_x^2(\vnab_\perp \cdot \bv_\perp)+\vnab\cdot \bff\,.
\eeqn
\ew
The left hand side of this expression obviously vanishes due to the incompressibility constraint \rf{divless}. Thus we are left with an equation relating the pressure $\cP$ to the velocity field $\bv$.  This equation can be further simplified by using
the condition \rf{divless} to replace $\vnab_\perp \cdot \bv_\perp$ with $-\pp_x v_x$ everywhere it appears.

We can then solve the resulting equation for $\cP$ in Fourier space; the result is:
\bew
\beqn
\cP=&&{iq_x\over (Kq_x^2+q_{_\perp}^2)}
\Bigg\{q_{_\perp}^2\left[\left([a-c]+\mu_{_\perp}q_{_\perp}^2+\mu_xq_x^2\right)v_x- f_x\right]
+\mathcal{F}_{\bq}\left[\Gamma_1v_x\partial_xv_x+
\right.\nonumber\\
&&\left.\Gamma_2 (\bv_{_\perp} \cdot\vnab_{_\perp})v_x
+\Gamma_3v_{_\perp}\partial_x v_{_\perp}+bv_x^3\right]
-i\bq\cdot\bff\Bigg\} \,,
\eeqn
\ew
 where we've defined
\begin{eqnarray}
\Gamma_1&\equiv&\lambda_1+\lambda_2-\lambda'_2+2 \left({\partial g\over\partial (v_x^2)}\right)_{\bv={\bf 0}}\,,\\
\Gamma_3&\equiv& 2\left({\partial g\over\partial (v_{_\perp}^2)}\right)_{\bv={\bf 0}}\,,
\end{eqnarray}
$\Gamma_2\equiv\lambda'_1$,
$\mu_x\equiv \mu_1+\mu_2-\mu'_2$,  and
$\mathcal{F}_{\bq}[g(\br,t)]\equiv{1\over (2\pi)^{d/2}}\int \dd^d \br\,e^{-i\bq\cdot\br}\,g(\br,t)$ represents the spatial Fourier transform of  $g(\br,t)$.

Inserting the resulting $\cP$ back into
(\ref{V_x1}), the equation becomes, in Fourier space,
\begin{widetext}
\begin{eqnarray}
\pp_t v_x(\bq,t) &=&-{1\over (Kq_x^2+q_{_\perp}^2)}
\left\{q_{_\perp}^2\left[\left(a+\mu_{_\perp}q_{_\perp}^2+\mu_xq_x^2\right)v_x- f_x\right]
+q_\perp^2\mathcal{F}_{\bq}\left[\Gamma_1v_x\partial_xv_x+
\right.\right.\nonumber\\
&&\left.\left.\Gamma_2 (\bv_{_\perp} \cdot\vnab_{_\perp})v_x
+\Gamma_3v_{_\perp}\partial_x v_{_\perp}+bv_x^3\right]
+q_x\left(wq_xv_x+K\bq_{_\perp}\cdot\bff_{_\perp}\right)\right\}
\,,\label{V_x2}
\end{eqnarray}
\end{widetext}
where we've further defined $w\equiv Kc$.
Note that we focus exclusively on the EOM of $v_x$ because it is the only soft mode at the critical transition.

\section{Mean-Field Theory}{\label{mft}}

We first consider the mean field theory for this system, which is simply its spatially uniform, steady state behavior in the absence of the noise $\bff$. In such a state, the equation of motion
 \rf{V_x1} reduces to a simple algebraic equation:
\beq
av_x+bv_x^3=0 \,.
\label{mfteom}
\eeq
For $a>0$, the only real solution of this equation is $v_x=0$ (we assume $b>0$ always). This is the disordered phase.

For $a<0$, we have two additional solutions:
\beq
v_x=\pm\sqrt{-{a\over b}} \,,
\label{mftsol}
\eeq
 which correspond to the ordered, broken symmetry state.

 We'll now consider fluctuations in both of these states, and at the critical point $a=0$ which separates them. We'll begin with the linear theory of those fluctuations.

 \section{Linear theory }{\label{lin}}

\subsection{Velocity Correlations in Fourier space }
\label{FTord}

In the linear approximation the fluctuations of the order parameter $v_x$   in the disordered state and at the critical point - that is, for $a\ge0$ - are readily
calculated. 
 One simply drops all of the  terms nonlinear in $v_x$ in the equation of motion \rf{V_x2}, and then Fourier transforms  in time. The result is a linear algebraic equation for the spatiotemporally Fourier transformed field $v_x(\bq, \omega)$, where $\omega$ is the Fourier frequency, which reads:
\begin{widetext}
\begin{eqnarray}
-i\omega v_x(\bq,\omega) &=&-{1\over (Kq_x^2+q_{_\perp}^2)}
\left\{q_{_\perp}^2\left[\left(a+\mu_{_\perp}q_{_\perp}^2+\mu_xq_x^2\right)v_x(\bq,\omega) - f_x(\bq,\omega) \right]
+q_x\left(wq_xv_x(\bq,\omega) +K\bq_{_\perp}\cdot\bff_{_\perp}(\bq,\omega) \right)\right\}
\,,\label{dislinV_x3}\nn\\
\end{eqnarray}
\end{widetext}
Solving this linear equation for $v_x(\bq,\omega)$ gives
\newpage
\bew
\beq
v_x(\bq,\omega)={q_\perp^2f_x+Kq_x\bqp\cdot\bff_{_\perp}(\bq,\omega)\over-i\omega(Kq_x^2+q_{_\perp}^2)+a q_{_\perp}^2+q_{_\perp}^2(\mu_xq_x^2+\mu_{_\perp}q_{_\perp}^2)+wq_x^2}
\,.
\label{vsol}
\eeq
\ew
 Autocorrelating the result \rf{vsol} with itself, and using the correlations of the random force $\bf$ as given by equation  (\ref{Noise_corr}), we find the spatio-temporally Fourier transformed velocity correlation function:
 \bew
\begin{eqnarray}
\label{eq:vxqo}
\langle v_x(\bq, \omega)v_x(\bq', \omega')\rangle
={2(D_x q_{_\perp}^4+D'_{_\perp}q_x^2q_{_\perp}^2)\delta(\bq+\bq')\delta(\omega+\omega')\over
\omega^2(Kq_x^2+q_{_\perp}^2)^2+\Big[a q_{_\perp}^2+q_{_\perp}^2(\mu_xq_x^2+\mu_{_\perp}q_{_\perp}^2)+wq_x^2\Big]^2}
\ ,
\end{eqnarray}
\ew
where  we've defined
$D'_{_\perp}\equiv D_{_\perp}K^2$. 
Integrating this over the frequencies $\omega$ and $\omega'$ gives the equal-time, spatially Fourier transformed correlation function:
%{and are given by}
\bew
\begin{eqnarray}
\label{eq:vx}
\langle v_x(\bq, t)v_x(\bq', t)\rangle=
{(D_x q_{_\perp}^4+D'_{_\perp}q_x^2q_{_\perp}^2)\delta(\bq+\bq')\over
(Kq_x^2+q_{_\perp}^2)\left[a q_{_\perp}^2+q_{_\perp}^2(\mu_xq_x^2+\mu_{_\perp}q_{_\perp}^2)+wq_x^2\right]}
\approx{(D_x q_{_\perp}^4+D'_{_\perp}q_x^2q_{_\perp}^2)\delta(\bq+\bq')\over
(Kq_x^2+q_{_\perp}^2)\left(a q_{_\perp}^2+\mu_{_\perp}q_{_\perp}^4+wq_x^2\right)}
\ ,
\label{vxETq}
\end{eqnarray}
\ew
Note that right at the critical point, where $a$ vanishes, as $\bq\to\mathbf 0$
the fluctuations diverge as
$1\over q_{\perp}^2$
for $q_x\lesssim q_{_\perp}^2$. For all other directions of $\bq$, this correlation function is
$O(1)$.  This implies anisotropic scaling:
$q_x\sim q_{_\perp}^2$, which we will show later to be modified by nonlinear effects in $d=2$, but not in $d=3$.  We have neglected
in the denominator  $\mu_xq_x^2q_{_\perp}^2$ with respected to $wq_x^2$ in the second approximation above, which is obviously justified in the limit of small $\bq$. We will therefore henceforth drop the $\mu_x$ term.

The above results apply in the disordered phase $a>0$. For $a<0$, i.e., in the ordered phase, we simply need to linearize our equation of motion not about $v_x=0$, but, rather, about the mean field solution (\ref{mftsol}). That is, we write
\beqn
v_x(\br,t)=v_0+\delta v_x(\br,t) 
\eeqn
with $v_0=\pm\sqrt{-{a\over b}}$, our mean field solution for $v_x$ from\\ equation \rf{mftsol} above, and then expand to linear order in $\delta v_x$. The result is
\begin{widetext}
\begin{eqnarray}
-i\omega \delta v_x(\bq,\omega) &=&-{1\over (Kq_x^2+q_{_\perp}^2)}
\left\{q_{_\perp}^2\left[\left(-2a+\mu_{_\perp}q_{_\perp}^2\right)\delta v_x(\bq,\omega) - f_x(\bq,\omega) \right]
+q_x\left(wq_x\delta v_x(\bq,\omega) +\bq_{_\perp}\cdot\bff'_{_\perp}(\bq,\omega) \right)\right\}
\,.
\nn\\
\label{V_x3}
\end{eqnarray}
\end{widetext}

We note that this is {\it identical} to the linearized equation of motion (\ref{dislinV_x3}) for $v_x$, with only the trivial changes  $v_x\to\delta v_x$ and $a\to-2a=2|a|$. Hence, we can simply make these substitutions in the results of the previous  subsection, to obtain
 \bew
\begin{eqnarray}
\label{eq:vx}
\langle \delta v_x(\bq, \omega)\delta v_x(\bq', \omega')\rangle=
{2(D_x q_{_\perp}^4+D'_{_\perp}q_x^2q_{_\perp}^2)\delta(\bq+\bq')\delta(\omega+\omega')\over
\omega^2(Kq_x^2+q_{_\perp}^2)^2+\left[2|a|q_{_\perp}^2
+q_{_\perp}^2(\mu_xq_x^2+\mu_{_\perp}q_{_\perp}^2)+wq_x^2\right]^2}
\ ,
\end{eqnarray}
\ew
for the connected spatio-temporally Fourier transformed correlation function,  and
\newpage
 \bew
\begin{eqnarray}
\langle \delta v_x(\bq, t)\delta v_x(\bq', t)\rangle=
{(D_x q_{_\perp}^4+D'_{_\perp}q_x^2q_{_\perp}^2)\delta(\bq+\bq')\over
(Kq_x^2+q_{_\perp}^2)\left(2|a| q_{_\perp}^2+\mu_{_\perp}q_{_\perp}^4+wq_x^2\right)}
\approx{(D_x q_{_\perp}^4+D'_{_\perp}q_x^2q_{_\perp}^2)\delta(\bq+\bq')\over
(Kq_x^2+q_{_\perp}^2)\left(2|a| q_{_\perp}^2+\mu_{_\perp}q_{_\perp}^4+wq_x^2\right)}
\ ,
\label{vxETq}
\end{eqnarray}
\ew
for the equal-time, spatially Fourier transformed correlation function.

\subsection{Velocity Correlations in real space and time  in the linear theory}

To obtain the spatio-temporal velocity correlation function, we will Fourier transform the above results back to real space and time. We will focus on the disordered phase exclusively in the following calculations. Once we have the results for the disordered phase, we can easily construct those for the ordered phase simply by replacing $a$ with $2|a|$,  utilizing  the striking similarity between (\ref{eq:vxqo}) and (\ref{eq:vx}).

Fourier transform the expression \rf{eq:vxqo} back to 
real space and time to obtain 
\bew
\beq
C({\bf r}, t)
=\int\frac{\dd^dq \, d\omega}{(2 \pi)^{d+1}}
{2(D_x q_{_\perp}^4+D'_{_\perp}q_x^2q_{_\perp}^2) \ee^{\ii (\bq \cdot \br-\omega t)}\over
\omega^2(Kq_x^2+q_{_\perp}^2)^2+\left(aq_{_\perp}^2+\mu_{_\perp}q_{_\perp}^4+wq_x^2\right)^2}\label{eq:vx2.0ut}
\ ,
\eeq
\ew
where $C({\bf r}, t)$ is defined by (\ref{cvvdef}).

 The correlation function behaves very differently
 in the ``critical regime" than in the ``non-critical regime. Here the ``critical regime" is defined as the regime in which  all of the following   three conditions are satisfied:
  \beq
 x\ll\xpa^{\rm lin} \sep r_{_\perp}\ll\xpe^{\rm lin}  \sep t\ll\tc^{\rm lin}\,,
 \label{crit reg lin}
 \eeq
where we've defined the correlation lengths and times for the linear theory:
\beqn
&&\xpe^{\rm lin}\equiv\sqrt{\mu_{_\perp}\over |a|} \sep \xpa^{\rm lin}\equiv\sqrt{w\over\mu_{_\perp}}(\xpe^{\rm lin})^2={\sqrt{w\mu_{_\perp}}\over |a|}\,,
\nn\\
&&\tc^{\rm lin}\equiv{1\over |a|}\,.
\label{corrlin}
\eeqn

The non-critical regime is everywhere else; i.e., where at least one of the conditions \rf{crit reg lin} is violated.

Note that the correlation lengths $\xpa^{\rm lin}$ and $\xpe^{\rm lin}$ both diverge  as $a\to0$. Indeed, 
if $a$ is a function of some experimental control parameter $p$, and $p_c$ is the value of $p$ satisfying $a(p_c)=0$, and if, as is the usual assumption in critical phenomena \cite{chaikin}, $a(p)$ is a smooth, analytic function of $p$ near $p_c$, then asymptotically close to $p_c$, we`ll have
\beq
a(p)\approx C(p-p_c) \,,
\label{acrit}
\eeq
with $C$ a non-universal (i.e., system-dependent), non-zero constant.
Inserting this into \rf{corrlin}, we see that both $\xpa^{\rm lin}$ and $\tc^{\rm lin}$ diverge like 
$|p-p_c|^{-1}$ as $p\to p_c$, while $\xpe^{\rm lin}$ diverges like $|p-p_c|^{-1/2}$.

We'll now determine the behavior of the spatial and temporal correlation functions, in the linear approximation, starting with the critical regime.

 \vspace{.2in}

\subsubsection{General scaling behavior of velocity correlations in the critical regime in the linear theory}

In the critical regime, where all conditions \rf{crit reg lin} are satisfied, the integral in (\ref{eq:vx2.0ut}) is dominated by $q_{_\perp}\gg1/\xpe^{\rm lin}$. To see this, note first that the integral over the inner regime $q_{_\perp}\ll 1/\xpe^{\rm lin}$, $q_x\ll1/\xpa^{\rm lin}$, $\omega\ll1/\tc^{\rm lin}$ converges. It has no contribution to the singular dependence on space $\br$ and time $t$, since the only dependence on space $\br$ and time $t$ comes from the exponential, which, in the inner regime, is practically 1, since $\bqp\cdot\brp$ and $q_xx$ are both $\ll1$ in the critical regime. So the singular spatial and temporal dependence can only comes from integrating over the outer regime where $a q_{_\perp}^2$ is negligible.
 In this limit, it is easy to see that the $a q_{_\perp}^2$ term in the denominator in (\ref{eq:vx2.0ut}) is negligible compared to the $\mu_{_\perp}q_{_\perp}^4$ term. We will therefore drop it in this regime.

In addition, as noted earlier, in the critical regime $q_x\sim q_{_\perp}^2\ll q_{_\perp}$; hence, we can drop the $Kq_x^2$ term in the denominator (\ref{eq:vx2.0ut})   relative to the $q_{_\perp}^2$ term. Likewise, we can drop the $D'_{_\perp}q_x^2q_{_\perp}^2$ term in the numerator relative to the $D_xq_{_\perp}^4$ term. Dropping these terms as well as the $a q_{_\perp}^2$ term in the denominator leaves us with
\beq
C({\bf r}, t)
=\int\frac{\dd^dq\,d\omega}{(2 \pi)^{d+1}}
{2D_x q_{_\perp}^4 \ee^{\ii (\bq \cdot \br-\omega t)}\over
\omega^2 q_{_\perp}^4+\left(\mu_{_\perp}q_{_\perp}^4+wq_x^2\right)^2}\label{eq:vx2.0}
\ .
\eeq
Making the change of variables of integration from $(\bqp, q_x, \omega)$ to $(\bQp, Q_x, \Omega)$ defined by
\beq
\bqp\equiv{\bQp\over \rp} \sep q_x\equiv {Q_xx_0^{\rm lin}\over \rp^2}
 \sep \omega\equiv{\Omega\mu_{_\perp}\over \rp^2} \,,
 \label{varchange}
 \eeq
we immediately find that the real space correlation function takes a scaling form:
\beq
C({\bf r}, t)=r_{_\perp}^{2\chi_{\rm lin}}F_{\rm lin}\left({x_0^{\rm lin}x\over r_{_\perp}^{\zeta_{\rm lin}}},{\mu_{_\perp} t\over r_{_\perp}^{z_{\rm lin}}}\right)
\label{eq:sfgeneral}
\eeq
with the crossover function $F_{\rm lin}(X,Y)$ given by
\bew
\beq
F_{\rm lin}(X,Y)
={2D_x\over\sqrt{\mu_\perp w}}\int\frac{\dd^dQ \, \dd\Omega}{(2 \pi)^{d+1}}
{ Q_{_\perp}^4 \ee^{\ii (\bQp\cdot\hat{\br}+Q_x X-\Omega Y)}\over
\Omega^2Q_{_\perp}^4+\left(Q_{_\perp}^4+Q_x^2\right)^2}\label{eq:sflin}
\,,
\eeq
\ew
and the scaling exponents in the linear theory given by
\beq
{\zeta_{\rm lin}}={z_{\rm lin}}=2 \sep \chi_{\rm lin}={1-d\over2} \,.
\label{linexp}
\eeq

In \rf{varchange}, we've defined the microscopic length
 \beq
 x_0^{\rm lin}\equiv\sqrt{\mu_{_\perp}\over w}
 \label{xodef}
 \eeq
 which goes to a finite constant as the transition is approached.

The scaling form (\ref{eq:sfgeneral}) implies (\ref{crit corr})  quoted in the introduction with
\beq
\chi=\chi_{\rm lin}\,, ~~z=z_{\rm lin}\,, ~~\zeta=\zeta_{\rm lin}\,,
\eeq
and the following connection between the non-universal non-critical microscopic lengths and time:
\beq
a_{_\perp}=\sqrt{a_x x_0^{\rm lin}}=\sqrt{\mu_{_\perp}\tau_0}\,.
\eeq

In three dimensions, all of the linear results for the {\it critical} regime just quoted continue to hold in the full theory. In two dimensions, they do not.

\subsubsection{Equal-time velocity correlations in the critical regime in the linear theory}

As we did in the previous subsection, we will, for the critical regime,  drop the $a q_{_\perp}^2$ term and 
the $Kq_x^2$ term in the denominator in (\ref{eq:vx2.0ut}) ,  and the $D'_{_\perp}q_x^2q_{_\perp}^2$ term in the numerator. 
Dropping these terms and setting $t=0$ leaves us with
\beq
C({\bf r}, 0)=\int\frac{\dd^dq\,d\omega}{(2 \pi)^{d+1}}
{2D_x q_{_\perp}^4 \ee^{\ii \bq \cdot \br}\over
\omega^2 q_{_\perp}^4+\left(\mu_{_\perp}q_{_\perp}^4+wq_x^2\right)^2}\label{eq:vx2.0et}
\ .
\eeq

This is an  analytically tractable integral; the result in $d=3$ is
\beqn
C({\bf r}, 0)
={D_x\over8\pi\mu_{_\perp}|x|}\exp\left(-{r_{_\perp}^2\over4x_0^{\rm lin}|x|}\right) \sep
(d=3)\,.
\nn\\
\label{rscorrd=3}
\eeqn
Note that this result agrees with the scaling form (\ref{eq:sfgeneral}), with the crossover function at equal time given by
\beq
F_{\rm lin}(X,0)={D_x\over8\pi\sqrt{\mu_\perp w}}\left[{\exp\left(-{1\over4|X|}\right)\over |X|}\right]\sep (d=3)  \,.
\label{3detcross}
\eeq

In $d=2$ in the critical regime, we find % \vspace{.2in}
\beq
C({\bf r}, 0)={D_x\over4(\mu_{_\perp}^3w)^{1/4}}\left[\exp\left(-{y^2\over4x_0^{\rm lin}|x|}\right)\over \sqrt{\pi|x|}\right]\,,
(d=2) \,.
\label{rscorrd=2}
\eeq
where we've taken the single Cartesian component of $\brp$ in $d=2$ to be $y$.

Note that this result also agrees with the scaling form (\ref{eq:sfgeneral}), with the crossover function at equal time given by
\beq
F_{\rm lin}(X,0)={D_x\over4\sqrt{\pi\mu_\perp w}}\left[{\exp\left(-{1\over4|X|}\right)\over \sqrt{|X|}}\right]\sep (d=2)  \,.
\label{2detcross}
\eeq

\subsubsection{Equal-time velocity correlations in the non-critical regime in the linear theory}\label{nccorrlin}

Outside the critical regime -i.e., when one or more of the conditions (\ref{crit reg lin}) is violated - we can ignore the  $\mu_\perp q_\perp^4$ term in the denominator of (\ref{eq:vx2.0ut}), 
To see this, let's consider this specific case $r_{_\perp}\gg\xpe^{\rm lin}$, $x=0$, $t=0$. In this case the integral is dominated by the regime $q_{_\perp}\ll \left(\xpe^{\rm lin}\right)^{-1}$, since, for $q_{_\perp}\gg\left(\xpe^{\rm lin}\right)^{-1}$, the magnitude of the integrand is very small, and the integral is canceled off by the oscillating exponential. Therefore, in the dominant regime,  $\mu_\perp q_\perp^4$ is negligible, since it is smaller than  $a q_\perp^2$. This argument applies to other cases outside the critical regime -i.e., $|x|\gg\xpa^{\rm lin}$, $r_{_\perp}=0$, $t=0$ or $|t|\gg\tc^{\rm lin}$, $x=0$, $r_{_\perp}=0$ - and leads to the same conclusion.

Ignoring the  $\mu_\perp q_\perp^4$ term in the denominator of (\ref{eq:vx2.0ut}), setting $t=0$,  and integrating over $\omega$ gives
\beq
C({\bf r}, 0)
=\int\frac{\dd^dq}{(2 \pi)^d}
{(D_x q_{_\perp}^4+D'_{_\perp}q_x^2q_{_\perp}^2) \ee^{\ii \bq \cdot \br}\over
(Kq_x^2+q_{_\perp}^2)(aq_{_\perp}^2+wq_x^2)}
\label{eq:vx2}
\ .
\eeq
In $d=3$, it is convenient to rewrite (\ref{eq:vx2}) as
\beq
C({\bf r}, 0)
=-\nabla_\perp^2\left(\int\frac{\dd^3q}{(2 \pi)^3}
{(D_x q_{_\perp}^2+D'_{_\perp}q_x^2) \ee^{\ii \bq \cdot \br}\over
(Kq_x^2+q_{_\perp}^2)(aq_{_\perp}^2+wq_x^2)}\right)
\label{Cvderiv}
\,.
\eeq
The integral in this expression is
\begin{widetext}
\beqn
g(\br)
=\int\frac{\dd^{2}q_{_\perp}}{(2 \pi)^{2}}\,e^{\ii\bqp\cdot\brp}\int_{-\infty}^\infty{\dd q_x\over2\pi}{(D_x q_{_\perp}^2+D'_{_\perp}q_x^2) \ee^{\ii q_x x}\over
(Kq_x^2+q_{_\perp}^2)(aq_{_\perp}^2+wq_x^2)}\,.
\label{f1}
\eeqn
\end{widetext}
The integral over $q_x$ can be evaluated by simple complex contour techniques, and gives
\beqn
&&
\int_{-\infty}^\infty{dq_x\over2\pi}{(D_x q_{_\perp}^2+D'_{_\perp}q_x^2) \ee^{\ii q_x x}\over
(Kq_x^2+q_{_\perp}^2)(aq_{_\perp}^2+wq_x^2)}\nonumber
\\
&=&
\frac{A_1\exp[-{q_\perp|x|\over\sqrt{K}}]
+B_1\exp[-q_\perp|x|\sqrt{\alpha_{\rm lin}}]}{2(aK-w)q_\perp}\,,~~~
\label{qxint}
\eeqn
 where we've defined the dimensionless parameter
\beqn
\alpha_{\rm lin}\equiv {a\over w}={\mu_{_\perp}\over w(\xpe^{\rm lin})^2}=\left(x_0^{\rm lin}\over \xpe^{\rm lin}\right)^2\,,
\label{alpha}
\eeqn
and the dimensionful constants
\beqn
A_1&\equiv& D_x\sqrt{K}-{D'_{_\perp}\over\sqrt{K}} \ ,
\\
B_1 &\equiv& -{D_x\over\sqrt{\alpha_{\rm lin}}}+D'_\perp\sqrt{\alpha_{\rm lin}} \,.
\label{ABdef}
\eeqn
In the second equality for $\alpha_{\rm lin}$, we have used our definition \rf{corrlin} of the correlation length $\xpe$.

Inserting \rf{qxint} into \rf{f1} gives
\bew
\beq
g(\br)={1\over2(aK-w)}\int\frac{\dd^{2}q_{_\perp}}{(2 \pi)^{2}}\,\left({1\over q_{_\perp}}\right) \left(A_1\exp\bigg[i\bqp\cdot\brp-{q_{_\perp}|x|\over\sqrt{K}}\bigg]+B_1\exp\bigg[i\bqp\cdot\brp-q_{_\perp}|x|\sqrt{\alpha_{\rm lin}}
\bigg]\right) \,.
\label{f2}
\eeq
\ew

Evaluating the integral of the $B_1$ term in polar coordinates gives
\bew
\beqn
\int\frac{\dd^{2}q}{(2 \pi)^{2}}\,\left({1\over q_\perp}\right)\exp\bigg[i\bqp\cdot\brp-q_\perp|x|\sqrt{\alpha_{\rm lin}}\,\bigg]&=&\int_0^{2\pi}{\dd\theta\over(2\pi)^2}\int_0^\infty \exp\bigg[-q_\perp\bigg(|x|\sqrt{\alpha_{\rm lin}}-i\rp\cos\theta\bigg)\bigg] dq_\perp
\nn\\
&=&\int_0^{2\pi}{\dd\theta\over(2\pi)^2}\left({|x|\sqrt{\alpha_{\rm lin}}+i\rp\cos\theta\over x^2 \alpha_{\rm lin}+\rp^2\cos^2\theta}\right)
 \,.
\label{b1}
\eeqn
\ew
The $\cos\theta$ factor in the numerator of this expression is odd under $\theta\to\theta+\pi$, while the denominator of the entire expression is even. Hence, that $\cos\theta$ factor in the numerator integrates to $0$, and can therefore be dropped, leaving us with
\beqn
\nn
&&\int\frac{\dd^{2}q}{(2 \pi)^{2}}\,\left({1\over q_\perp}\right)\exp\bigg[i\bqp\cdot\brp-q_\perp|x|\sqrt{\alpha_{\rm lin}}\,\bigg]
\\
&=&{|x|\sqrt{\alpha_{\rm lin}}\over(2\pi)^2}\int_0^{2\pi}{\dd\theta\over x^2 \alpha_{\rm lin}+\rp^2\cos^2\theta}
 \,.
\label{b2}
\eeqn
The final angular integral in this expression is elementary, and gives 
\beqn
\nn
&&
\int\frac{\dd^{2}q}{(2 \pi)^{2}}\,\left({1\over q_\perp}\right)\exp\bigg[i\bqp\cdot\brp-q_\perp|x|\sqrt{\alpha_{\rm lin}}\,\bigg]
\\
&=&{1\over2\pi\sqrt{x^2 \alpha_{\rm lin}+\rp^2}} 
 \,.
\label{bfin}
\eeqn
The $A_1$ term can be evaluated in precisely the same way; the result is an identical expression with the replacement $\al_{\rm lin}\to{1\over K}$.

Using these results in our expression \rf{f2} gives, in $d=3$,
\beq
g(\br)={1\over4\pi(aK-w)}\left[{A_1\over\sqrt{{x^2\over K}+\rp^2}}
+{B_1\over\sqrt{x^2 \alpha_{\rm lin}+\rp^2}} 
\right] \,.
\eeq
Inserting this into \rf{Cvderiv} gives
 \begin{eqnarray}
C({\bf r}, 0)={f_{_{3D}}(\theta)\over r^3}  \sep d=3\,
\label{eq:vx3}
\end{eqnarray}
where $\theta=\tan^{-1}\left({r_{\perp}\over |x|}\right)$ is the polar angle of $\br$ from the $x$-axis,
and 
\begin{widetext}
\beqn
f_{_{3D}}(\theta)&\equiv&
{
\xpe^{\rm lin}\over 4\pi\sqrt{\mup w}\left(1-{K\alpha_{\rm lin}}\right)
}
\left[
\left(D_x-D'_{_\perp}\alpha_{\rm lin}\right){(2\alpha_{\rm lin}\cos^2\theta-\sin^2\theta)\over(\sin^2\theta+\alpha_{\rm lin}\cos^2\theta)^{5/2}}\right.\nonumber\\
&&\left.
-K(KD_x-D'_{_\perp})\sqrt{\alpha_{\rm lin}}{(2\cos^2\theta-K\sin^2\theta)\over(K\sin^2\theta+\cos^2\theta)^{5/2}}
\right]\,.
\label{f3d}
%\nn\\
\eeqn
\end{widetext}

Near the critical point, since $\xpe^{\rm lin}\gg x_0^{\rm lin}$  and hence $\alpha_{\rm lin}\ll 1$, $f_{_{3D}}(\theta)$ reduces to
\beqn
f_{_{3D}}(\theta)=
{D_x\xpe^{\rm lin}\left(2\alpha_{\rm lin}\cos^2\theta-\sin^2\theta\right)\over 4\pi\sqrt{\mup w}(\sin^2\theta+\alpha_{\rm lin}\cos^2\theta)^{5/2}}\,.~~~~~
\eeqn
Note the result (\ref{eq:vx3}) for the equal-time correlation function in the non-critical region and that of (\ref{rscorrd=3}) in the critical region
connect at the crossover [i.e. $r_{_\perp}\sim\xpe^{\rm lin}, |x|\sim(\xpe^{\rm lin})^2 /x_0^{\rm lin}=\xpa^{\rm lin}$], where both scale as $(\xpe^{\rm lin})^{-2}$.

In $d=2$, we`ll consider first the case $r_{_\perp}>0$ (note that $r_{_\perp}$ is a scalar in $d=2$), and obtain the value of $C(\br,0)$ for $\rp<0$ by using the obvious fact that 
$C_{vv}(\br,0)$
is an even function of $\br$.

We begin by rewriting \rf{eq:vx2} as 
\beqn
\nn
C({\bf r}, 0)
&=&-\ii{\pp\over \pp \rp}\left(\int\frac{\dd^2q}{(2 \pi)^2}
{(D_x q_{_\perp}^2+D'_{_\perp}q_x^2) q_{_\perp}\ee^{\ii \bq \cdot \br}\over
(Kq_x^2+q_{_\perp}^2)(aq_{_\perp}^2+wq_x^2)}\right)
\\
&\equiv&-\ii{\pp h(\br)\over \pp \rp}
\label{Cvderiv2d}
\,,
\eeqn
where we`ve defined
\beq
h({\bf r})
\equiv\left(\int\frac{\dd^2q}{(2 \pi)^2}
{(D_x q_{_\perp}^2+D'_{_\perp}q_x^2) q_{_\perp}\ee^{\ii \bq \cdot \br}\over
(Kq_x^2+q_{_\perp}^2)(aq_{_\perp}^2+wq_x^2)}\right) \
\label{h1}
\,.
\eeq
Doing the integral over $q_{_\perp}$ by contours gives
\beqn
\nn
h({\bf r})
&=&{\ii\over4\pi}\int_{-\infty}^\infty \dd q_x \bigg[A_2\exp\bigg(\ii q_xx-\sqrt{K}|q_x|\rp\bigg)
\\
&&+B_2\exp\bigg(\ii q_xx-|q_x|\rp/\sqrt{\alpha_{\rm lin}}\bigg)\bigg] \,,
\label{h2}
\eeqn
where we've defined
\beqn
A_2\equiv {D'_\perp-KD_x\over (w-aK)} \sep B \equiv {{D_x\over\alpha_{\rm lin}}-D'_\perp\over (w-aK)} \,.
\label{AB2def}
\eeqn
The integral from $0$ to $\infty$ in this expression is the complex conjugate of the integral from $-\infty$ to $0$, and both are elementary. We thereby obtain
\beq
h({\bf r})
={\ii\rp\over2\pi}\bigg[{A_2\sqrt{K}\over  K\rp^2+x^2}+{B_2\sqrt{\alpha_{\rm lin}}\over \rp^2+\alpha_{\rm lin} x^2
}\bigg] \,.
\label{hfin}
\eeq
Inserting this into \rf{Cvderiv2d} and taking the derivative there gives
 \begin{eqnarray}
C({\bf r}, 0)={f_{_{2D}}(\theta)\over r^2}
\label{eq:vx4}
\end{eqnarray}
where
\begin{widetext}
\beqn
f_{_{2D}}(\theta)\equiv
{
\xpe^{\rm lin}\over 2\pi\sqrt{\mu_{_\perp} w}\left(1-K\alpha_{\rm lin}\right)
}
\left[\left(D_x-D'_{_\perp}\alpha_{\rm lin}
\right)
{(\alpha_{\rm lin}\cos^2\theta-\sin^2\theta)\over(\sin^2\theta+\alpha_{\rm lin}\cos^2\theta)^2}
-
\sqrt{K\alpha_{\rm lin}}\left(KD_x-D'_{_\perp}\right){(\cos^2\theta-K\sin^2\theta)\over(K\sin^2\theta+\cos^2\theta)^2}
\right]
\,,
\nn\\
\label{f_2D}
\eeqn
\end{widetext}
which, in the large-$\xpe^{\rm lin}$ limit (which is also the small $\alpha_{\rm lin}$ limit), reduces to 
\beqn
f_{_{2D}}(\theta)\equiv
{
\xpe^{\rm lin} D_x\over 2\pi\sqrt{\mu_{_\perp} w}
}
{(\alpha_{\rm lin}\cos^2\theta-\sin^2\theta)\over(\sin^2\theta+\alpha_{\rm lin}\cos^2\theta)^2}
\,.
\label{f2d}
\eeqn

\vspace{.2in}

Also note that the results  (\ref{eq:vx4}) for the equal-time correlation function in the non-critical region
and (\ref{rscorrd=2}) in the critical region
connect smoothly at the crossover [i.e. $r_{_\perp}\sim\xpe^{\rm lin}, |x|\sim(\xpe^{\rm lin})^2/x_0^{\rm lin}=\xpa^{\rm lin}$], where both scale as $(\xpe^{\rm lin})^{-1}$.

\subsubsection{Equal-position velocity correlations in the critical and non-critical regimes}
Setting $\br={\mathbf 0}$ in (\ref{eq:vx2.0ut}) and integrating over $\omega$,  we get
\bew
\beqn
&&C({\bf 0}, t)
=\int\frac{\dd^dq}{(2 \pi)^d}
{(D_x q_{_\perp}^4+D'_{_\perp}q_x^2q_{_\perp}^2)\exp\left[-{aq_{_\perp}^2+\mu_\perp q_\perp^4 +wq_x^2\over Kq_x^2+q_{_\perp}^2}|t|\right] \over
(Kq_x^2+q_{_\perp}^2)(aq_{_\perp}^2+\mu_\perp q_\perp^4+wq_x^2)}\label{eqtime_2v2}
\ .~~~
\eeqn
\ew

By inspecting the argument of the exponential, and the denominator, of 
the integral in \rf{eqtime_2v2}, we see that integral  is clearly dominated by values of $q_x$ such that 
\beq
wq_x^2\lesssim aq_{_\perp}^2+\mu_{_\perp} q_{_\perp}^4 \,.
\label{qxdom}
\eeq
If we are either at the critical point, where $a=0$, or near it, so that $a/w\ll1$, and we consider the limit of small $\bqp$ (specifically $q_{_\perp}\ll\sqrt{w\over\mu_{_\perp}}$, which is clearly the dominant regime in the integral \rf{eqtime_2v2} at large times), then we have
\beq
{q_x\over q_{_\perp}}\lesssim\sqrt{{a\over w}+\left({\mu_{_\perp}\over w}\right) q_{_\perp}^2}\ll1 \,,
\label{qxdom2}
\eeq
In light of this, we can drop the $D^\prime q_x^2q_{_\perp}^2$ in the numerator, and the $Kq_x^2$ terms everywhere they appear in \rf{eqtime_2v2}. Doing so reduces \rf{eqtime_2v2} to 
\beqn
\nn
&&C({\bf 0}, t)\equiv C_{ep}(t)
\\
&=&\int\frac{\dd^dq}{(2 \pi)^d}
{D_x q_{_\perp}^2\exp\left[-(a+\mu_{_\perp} q_{_\perp}^2 +w{q_x^2\over q_{_\perp}^2})|t|\right] \over
(aq_{_\perp}^2+\mu_{_\perp} q_{_\perp}^4+wq_x^2)}\nn\\
\label{eqtime_2v3}
\ .~~~
\eeqn
Differentiating both sides of this with respect to $|t|$ gives
\bew
\beq
{\dd C_{ep}(t)\over \dd |t|}=-D_xe^{-a|t|}\int\frac{\dd^{d-1}q_\perp}{(2 \pi)^{d-1}}e^{-\mu_\perp q_\perp^2|t|}\int_{-\infty}^\infty\frac{\dd q_x}{2 \pi}
\exp\left[-\left({wq_x^2\over q_\perp^2}\right)|t|\right] \,.
\eeq
\ew
Performing the Gaussian integral over $q_x$ gives
\beq
{\dd C_{ep}(t)\over \dd |t|}=-{D_xe^{-a|t|}\over2\sqrt{\pi w|t|}}\int\frac{\dd^{d-1}q_\perp}{(2 \pi)^{d-1}}q_\perp e^{-\mu_\perp q_\perp^2|t|} \,.
\label{dtcgen}
\eeq
 
 In $d=3$, we can evaluate  the $\int\dd^2q_\perp$ in polar coordinates. We obtain, after the trivial angular integral,
\beq
{\dd C_{ep}(t)\over \dd |t|}=-{D_xe^{-a|t|}\over4\sqrt{\pi ^3 w|t|}}\int_0^\infty \dd q_\perp \, q_\perp^2  e^{-\mu_\perp q_\perp^2|t|} \,.
%\sep d=3 \,.
\eeq
The final $\int dq_\perp$ is also a straightforward Gaussian integral, and leaves us with our final result for the time derivative of $C_{ep}(t)$:
\beq
{\dd C_{ep}(t)\over \dd |t|}=-\left({D_x\over16\pi\sqrt{w\mu_\perp^3}}\right)\left({e^{-a|t|}\over t^2}\right) \,.
%\sep d=3\,.
\eeq
Integrating this equation gives
\beq
C_{ep}(t)=\left({D_x\over16\pi\sqrt{w\mu_\perp^3}}\right)\int_{|t|}^\infty \dd u\, \left( {e^{-au}\over u^2}\right) + C_\infty \,.
%\sep d=3 \,.
\label{Cet1}
\eeq
The constant of integration $C_\infty$ is readily seen to be zero, by noting that $C_{ep}(t\to\infty)\to0$. Thus we have
\beq
C_{ep}(t)=\left({D_x\over16\pi\sqrt{w\mu_\perp^3}}\right)\int_{|t|}^\infty \dd u\, \left( {e^{-au}\over u^2}\right) \,.
%\sep d=3 \,.
\label{Cet2}
\eeq
Changing variable of integration from $u$ to $y=au$, and then integrating by parts,   gives 
\beqn
\nn
C_{ep}(t)&=&\left({D_x\over16\pi\sqrt{w\mu_\perp^3}}\right)\left[{e^{-a|t|}\over|t|}-a\int_{a|t|}^\infty \dd y\, \left( {e^{-y}\over y}\right) \right] 
\\
&=&\left({D_x\over16\pi\sqrt{w\mu_\perp^3}}\right)\left[{e^{-a|t|}\over|t|}+a{\rm Ei}(-a|t|) \right] \,,
%\sep d=3\,
\label{Cet3}
\eeqn
where in the final equality we have used the definition
\beq
{\rm Ei}(x)\equiv -\int_{-x}^\infty \dd y\,\left( {e^{-y}\over y}\right)
\label{eidef}
\eeq
of the exponential integral function ${\rm Ei}(x)$.

Using the well-known asymptotic expansions of this function for large and small argument, we see that in the critical regime $a|t|\ll1$, we have
\beq
C_{ep}(t)=\left({D_x\over16\pi\sqrt{w\mu_\perp^3}}\right)\left({1\over|t|} \right) 
%\sep d=3 \sep {\rm critical} \, {\rm regime}  \,\, a|t|\ll1 \,.
\label{Cet3dcrit}
\eeq
which is the result we claimed in the introduction in equation \rf{Cet3}.

In the opposite limit $a|t|\gg1$  (i.e., in the non-critical regime), we have
\beq
C_{ep}(t) =\left({D_x\over16\pi\sqrt{w\mu_\perp^3}}\right)\left[{e^{-a|t|}\over at^2} \right] \,.
%\sep d=3 \sep {\rm non-critical} \, {\rm regime}  \,\, a|t|\gg1 \,.
\label{Cet3dnoncrit}
\eeq

We now turn to the equal-position correlation function in $d=2$, \rf{dtcgen} implies
\beq
{\dd C_{ep}(t)\over \dd |t|}=-{D_xe^{-a|t|}\over\sqrt{\pi w|t|}}\int_0^\infty\frac{\dd q_\perp}{2 \pi}q_\perp e^{-\mu_\perp q_\perp^2|t|} \,.
%\sep d=2\,.
\label{dtcd=2}
\eeq
The integral over $q_\perp$ is elementary; we obtain:
\beq
{\dd C_{ep}(t)\over \dd|t|}=-\left({D_xe^{-a|t|}\over4\mu_\perp\sqrt{\pi^3w|t|^3}}\right)\,. %\sep d=2\,.
\eeq
Integrating this and using the fact that $C_{ep}(t\to\infty)\to0$ to fix a constant of integration, we obtain 
\beq
C_{ep}(t)=\left({D_x\over4\mu_\perp}\right)\sqrt{a\over\pi^3w}\,\,\Gamma\left( -{1\over2} , a|t|\right)\,,
%\sep d=2\,,
\eeq

where
\beq
\Gamma( w, u)\equiv\int_u^\infty e^{-v}v^{w-1}\,\dd v
\label{incgamdef}
\eeq
is the incomplete Gamma function.

Using the well-known asymptotic expansions of this function for large and small argument, we see that in the critical regime $a|t|\ll1$, we have
\beq
C_{ep}(t)=+\left({D_x\over2\mu_\perp\sqrt{\pi^3w|t|}}\right)\,.
%\sep d=2  \sep {\rm critical} \, {\rm regime}  \,\, a|t|\ll1 \,.
\label{Cet2dcrit}
\eeq

 In the opposite limit $a|t|\gg1$ (i.e., in the non-critical regime) we have
 \beqn
C_{ep}(t)&=&{D_xe^{-a|t|}\over4a\mu_\perp\sqrt{\pi^3w|t|^3}}
\nn\\
&=&
{D_x\over 4\sqrt{\pi^3\mu_{_\perp}w}\xi_{_\perp}^{\rm lin}}
\left(|t|\over\tau_{\rm corr}^{\rm lin}\right)^{-{3\over 2}}
\exp\left(-|t|\over \tau_{\rm corr}^{\rm lin}\right)\,,~~~~~~~~
%\sep d=2  \sep {\rm non-critical} \, {\rm regime}  \,\, a|t|\gg1 \,, 
\label{Cet2dnoncrit}
\eeqn
where $\xi_{_\perp}^{\rm lin}$ and $\tau_{\rm corr}^{\rm lin}$ are given by \rf{corrlin}.

\section{Identification of the relevant non-linearities}{\label{id}}

Knowing the form (\ref{eq:vx2.0}) of the correlation function at the critical point, we now perform the standard power counting procedure on the model equation (\ref{V_x2}). We rescale time $t$,
space $(x,\brp)$, and the field $v_x$ as
\begin{eqnarray}
&&t\mapsto te^{z\ell} \ , \ \ x\mapsto xe^{\zeta\ell}\ ,
\ \ \br_{_\perp}\mapsto\br_{_\perp}e^{\ell}\,,\\
&&v_x\mapsto v_xe^{\chi\ell}\ ,\ \ \bv_{_\perp}\mapsto\bv_{_\perp} e^{(\chi-\zeta+1)\ell}\,,
\end{eqnarray}
and will choose the dynamical exponent $z$, the anisotropy exponent $\zeta$, and the ``roughness" exponent $\chi$ to keep the  coefficients $\mu_{_\perp}$, $w$, and the random force strength $D_x$ that appear in the correlation function  (\ref{eq:vx2.0})  {\it at the critical point} unchanged.  Performing the rescaling, we find
\begin{eqnarray}
&&\mu_{_\perp}\mapsto \mu_{_\perp}e^{(z-2)\ell}\ ,\ \ w\mapsto we^{(z+2-2 \zeta)\ell}\ ,\
\\
&&D_x\mapsto D_xe^{(z-2\chi-d+1-\zeta)\ell}\,.
\end{eqnarray}
Choosing $z$, $\zeta$ and $\chi$ such that $\mu_{_\perp}$, $w$ and $D_x$ are fixed upon the rescaling gives
\begin{eqnarray}
\label{eq:MFexp}
z=2\ ,\ \ \zeta=2\ ,\ \ \chi={1-d\over 2}\,.
\end{eqnarray}
With this choice of the rescaling exponents, the coefficients of the various
terms in (\ref{V_x2}) and $D'_{_\perp}$, the noise strength of the redefined
noise $\bff'_{_\perp}$, scale as the following:
\begin{eqnarray}
&&a\mapsto ae^{2\ell}\,,~b\mapsto be^{(3-d)\ell}\,,\\
&&K\mapsto Ke^{-2\ell}\,,~\mu_x\mapsto be^{-2\ell}\,,\\
&&\Gamma_1\mapsto \Gamma_1e^{\left({1-d\over 2}\right)\ell}\,,
~\Gamma_2\mapsto \Gamma_2e^{\left({1-d\over 2}\right)\ell}\,,\\
&&\Gamma_3\mapsto \Gamma_3e^{-\left({3+d\over 2}\right)\ell}\,,
D'_{_\perp}\mapsto D'_{_\perp}e^{-2\ell}\,.
\end{eqnarray}
This shows that only $a$ and $b$ are non-vanishing in
the dimensions of physical interest (i.e., $1<d\leq 3$) in
the limit $\ell\to\infty$. In contrast,
all the $\Gamma_i$'s, $K$, $\mu_x$,
and  $D'_{_\perp}$ vanish in those
dimensions in the same limit.  This implies all the
terms associated with these vanishing parameters in (\ref{V_x2}) are
irrelevant, and hence negligible in the hydrodynamic
limit.
Dropping all of these irrelevant terms leads to the following
 simplified hydrodynamic model,  which, although simplified, contains all terms relevant for
the critical behavior of the system for small $a$:
\begin{eqnarray}
\pp_t v_x(\bq,t)
&=&-\left(a+\mu_{_\perp}q_{_\perp}^2+w{q_x^2\over q_{_\perp}^2}\right)v_x(\bq,t)
\nonumber
\\
&&
-b\mathcal{F}_{\bq}\left[v_x^3(\br,t)\right]+f_x(\bq,t)
\,.\label{V_x3}
\end{eqnarray}

\section{Mapping onto the Ising model with dipolar interactions}{\label{map}}

By inspection, we can see that (\ref{V_x3}) can be re-written in the form of a functional derivative:
\beq
\pp_t v_x = -\Gamma\frac{\delta H}{\delta v_x} +f_x
\ ,
\label{TDGL}
\eeq
where
\beqn
\label{eq:H}
&&H = \frac{1}{2}\int_\bq  \left[m+c_sq_{_\perp}^2+c_d{q_x^2\over q_{_\perp}^2}\right]v_x(\bq)v_x(-\bq)
\\ \nonumber
&&+\frac{u}{4}\frac{1}{(2\pi)^d}
\int_{\bq,\bq_{1,2}}v_x(\bq)v_x(\bq_1)v_x(\bq_2)v_x(-\bq-\bq_1-\bq_2)
\ ,
\eeqn
and $\int_\bq \equiv \int \dd^d\bq$.
This is readily seen to recover \rf{V_x3}
with
\beq
a=\Gamma m \sep \mu_\perp=\Gamma c_s \sep w=\Gamma c_d \sep b=\Gamma u \,.
\label{eqcon}
\eeq
To complete the analogy with an equilibrium system, we note that the fluctuation dissipation theorem \cite{FDT}  requires that the noise correlation and the kinetic coefficient  in \rf{TDGL} be related by
\begin{eqnarray}
\langle f_x(\br, t)f_x(\br',t')\rangle
=2\Gamma\kbt
\delta^d(\br-\br')\delta(t-t')\,.~~~~~
\label{Noise_corr_eq}
\end{eqnarray}
This condition  ensures that the equation of motion \rf{TDGL} implies that at long times the steady-state probability distribution $P(\{v_x(\br,t\})$  of a specified spatial configuration of  the field $v_x(\br,t)$ at any fixed time $t$ is time-independent, and given simply by the Boltzmann weight
\beq
P(\{v_x(\br,t\})={e^{-\beta H[\{v_x(\br,t\} ]}\over Z} \,,
\label{bolt}
\eeq
with $\beta\equiv1/\kbt$
associated with the Hamiltonian \rf{eq:H}, where $Z$ is the partition function for $H$. 

Comparing \rf{Noise_corr_eq} with our earlier expression \rf{Noise_corr} for the noise correlations, we see that the effective temperature of our system is given by
\beq
\kbt={D_x\over\Gamma} \,.
\label{Teff}
\eeq

When irrelevant terms are restored, the 
dynamical model \rf{TDGL} is a simple, purely relaxational model that relaxes back to the equilibrium steady state of the Hamiltonian  \rf{eq:H}, and its critical dynamics have also been studied \cite{folk_zpb77} using the dynamical renormalization group (DRG).
 We will review this analysis and its results in the next section.

 To perform the RG analysis of the next section, it is convenient to separate the Hamiltonian $H$ into harmonic and anharmonic parts $H_h$ and $H_a$ respectively. That is, we write
\beq
H\equiv H_h[\{v_x(\br,t\}]+H_a[\{v_x(\br,t\}]
\label{hsep}
\eeq
with
\beq
H_h\equiv \frac{1}{2}\int_\bq  \left[m+c_sq_{_\perp}^2+c_d{q_x^2\over q_{_\perp}^2}\right]v_x(\bq)v_x(-\bq)
\label{hhdef}
\eeq
and 
\beq
H_a\equiv \frac{u}{4}\frac{1}{(2\pi)^d}
\int_{\bq,\bq_{1,2}}v_x(\bq)v_x(\bq_1)v_x(\bq_2)v_x(-\bq-\bq_1-\bq_2) \,.
\label{hadef}
\eeq

 \section{Renormalization Group analysis}{\label{RGsec}

  To deal with the effects of the $u$ non-linearity in our equation of motion \rf{TDGL}, we can use a variety of renormalization group approaches. If we are only interested in the {\it statics} - that is, the equal-time correlation functions - then we can use the static renormalization group developed by Kadanoff, Wilson, and Fisher \cite{KWF}, applied directly to the Boltzmann weight associated with the Hamiltonian \rf{eq:H}. This was the approach taken by \cite{aharony_ising73} and \cite{brezin_prb76}. To determine  
time dependent correlations, we can supplement the RG results from the static RG with a dynamical renormalization group analysis, as was done by \cite{folk_zpb77}. 

We will now review these two approaches here. Our approach, and, therefore, our recursion relations, differ  slightly  in some technical details from those of \cite{aharony_ising73, brezin_prb76, folk_zpb77} (particularly in our choice of the Brillouin zone shape). Our results for physically observable quantities are, of course,  exactly the same as theirs, since there is only one correct answer!

In the next subsection, we will explain the mechanics of the renormalization group, and present the recursion relations.

\subsection{The renormalization group (RG)  approach and recursion relations in $d\le3$}{\label{rr}}

\subsubsection{The static RG}

The {\it static} RG \cite{KWF} starts by ignoring the time dependence of the field 
$v_x(\br,t)$, because the steady-state probability distribution $P(\{v_x(\br,t\})$  of a specified spatial configuration of  the field $v_x(\br,t)$ at any fixed time $t$ is time-independent, and given simply by the Boltzmann weight
\beq
P(\{v_x(\br,t\})={e^{-\beta H[\{v_x(\br,t\} ]}\over Z} \,,
\label{bolt}
\eeq
with $\beta\equiv1/\kbt$, and $\kbt$ given by equation \rf{Teff}, 
associated with the Hamiltonian \rf{eq:H}, where $Z$ is the partition function for $H$. Since this is independent of time, we will, for the remainder of this discussion of the static RG, drop the time argument of $v_x(\br,t)$.

The static RG now proceeds in three steps. First, we separate the field
$v_x(\br)$ into ``fast'' and ``slow''
components $v_x^>(\br)$ and $v_x^<(\br)$, where the ``fast''
component $v_x^>(\br)$ evolves rapidly in space, while the ``slow'' component  $v_x^<(\br)$ evolves slowly in space. More precisely, 
we write 
\beq
v_x(\br)=v_x^<(\br)+v_x^>(\br) \,,
\label{fastlowsp}
\eeq
where the fast component $v_x^>(\br)$ only has support in the infinitely long $d$-dimensional hypercylinder ``shell'' of Fourier
space $b^{-1} \Lambda \le |\bq_\perp| \le \Lambda$,  $-\infty<q_x<\infty$, where $\Lambda$ is an
``ultra-violet cutoff", while the ``slow'' component $v_x^<(\br)$
has support in the ``core'' $0 \le |\bq_\perp| \le b^{-1} \Lambda$,  $-\infty<q_x<\infty$ (Hence the superscripts $>$ and $<$ for the fast and slow components respectively). Here
$\Lambda$ is of order the inverse of some microscopic length $a$ (e.g., the inter-bird distance). The rescaling factor
$b$ is $>1$, but otherwise arbitrary.  Later, to obtain {\it
  differential} recursion relations, we will take $b \equiv e^{\dd\ell}$ to
be close to $1$; that is, we`ll take  $\dd\ell \ll 1$.

The second step of the static RG is to  derive an ``intermediate'' Hamiltonian $H_I(\{v^<_x(\br)\})$ for the
slow degrees of freedom $v_x^<$ by integrating the Boltzmann weight
$Z^{-1}e^{-\beta H[\{v_x^<(\br), v_x^>(\br)\}]}$ over the ``fast''
degrees of freedom $v_x^>(\br)$.  That is,
we write
 \bew
 \beq
Z_I^{-1} e^{-\beta H_I[\{v^<_x(\br)\}]}=Z^{-1}\int
\prod_{\bq} \dd v_x^>(\bq)\; e^{-\beta H[\{v_x^<(\br),v_x^>(\br)\}]}
= Z_0 Z^{-1}e^{-\beta H_h[\{v_x^<(\br)\}]}
\big\langle e^{-\beta
  H_a[\{v_x(\br),v_x^>(\br)\}]}\big\rangle_0^>
\,,
\eeq
\ew
where $\int \prod_{\bq} dv_x^>(\bq)$ denotes an integral over {\it
  only} the fast degrees of freedom $v_x^>(\bq)$, and the symbol
$\langle\ldots \rangle_0^>$ denotes an average of $\ldots$ over those
fast modes using {\it only} the Boltzmann weight
$ Z_0^{-1} e^{-\beta H_h[v_x^>(\br)]}$ for the purely quadratic part
of the Hamiltonian.  Since this Hamiltonian is quadratic, these
averages are straightforward to evaluate in perturbation theory (which amounts to expanding the exponential in $\big\langle e^{-\beta
  H_a[\{v_x(\br),v_x^>(\br)\}]}\big\rangle_0^>$, being averages over a purely
Gaussian distribution.

As usual in DRG calculations \cite{fns}, we need to make approximations to perform the averaging step.

 The third and final step of the static RG is to  rescale time, lengths, and  fields as
\beqn
x\to xb^\zeta \sep \brp\to \brp b %\,,\nn\\
\sep v_x\to v_xb^{\chi} \,,
\label{rescale}
\eeqn
to restore  the Brillouin zone to its original size.  The ``anisotropy" exponent $\zeta$ and the ``roughness" exponent $\chi$ are at this point arbitrary. We will later choose them to produce fixed points of the RG.

The RG now proceeds by iterating this process. The result can be summarized by {\it differential} recursion relations in the following (by now very standard) manner:
as already mentioned, we choose $b=1+\dd\ell$ with $\dd\ell$ differential. Instead of keeping track of the number $n$ of iterations of the renormalization group, we introduce a ``renormalization group time'' $\ell$ defined as $\ell\equiv n\dd\ell$. 

The resultant renormalized Hamiltonian now becomes a function of the RG time $\ell$. Since the symmetry of our system remains the same under this process, the {\it form} of the Hamiltonian remains that of equation \rf{eq:H}, but with the original (or ``bare`` parameters $m$, $c_s$, $c_d$, and $u$ replaced with ``renormalized" values $m(\ell)$, $c_s(\ell)$, $c_d(\ell)$, and $u(\ell)$ 
which depend on the RG time $\ell$. By evaluating the changes in these parameters for $b=1+\dd\ell$, we (and references \cite{aharony_ising73} and \cite{brezin_prb76} long before us) can derive
differential equations governing the evolution of these renormalized parameters with $\ell$.

As mentioned earlier, the second step (i.e., the averaging over the fast modes $v_x^>$) can only be done perturbatively in the non-linearity $u(\ell)$. We will also expand these recursion relations to linear order in $m$, since we are interested in the transition, which occurs near $m=0$. One final approximation is to evaluate the graphical corrections in $d=3$, which we justify by focusing on spatial dimensions near $3$. The resultant recursion relations read \cite{aharony_ising73, brezin_prb76} 
\beqn
 {\dd m\over \dd\ell} &=&(2\chi+\zeta+d-1)m+ 6gc_s\Lambda^2-3gm +O(gm^2, g^2)\nn\\ \label{mrr} \\
 {\dd c_s\over \dd\ell}&=&\left(2\chi+\zeta+d-3+{4g^2\over3}+O(mg^2, g^3)\right)c_s
 \label{csrr}\\
 {\dd c_d\over \dd\ell}&=&(2\chi-\zeta+d+1)c_d
 \label{cdrr}\\
 {\dd u\over \dd\ell} &=&\bigg(4\chi+\zeta+d-1 - 9g+O(g^2, mg)\bigg)u 
 \label{rrs fin}
\eeqn
where we`ve defined the dimensionless coupling
\beq
g\equiv{S_{d-1}\Lambda^{d-3}u\kbt\over4(2\pi)^{d-1}}\sqrt{c_s^3c_d} \,.
\label{gdef}
\eeq
 As noted by \cite{aharony_ising73, brezin_prb76, folk_zpb77}, the recursion relation \rf{cdrr} for $c_d$ is {\it exact}, to all orders in $u$ and $m$, since $c_d$ is the coefficient of 
a term that is non-analytic in wavevector $\bq$.

It is straightforward to construct the recursion relation for the dimensionless parameter $g$ itself from the above recursion relations. This is most easily done by considering $\ln{g}=\ln{u}-{3\over2}\ln{c_s}-{1\over2}\ln{c_d}+{\rm constant}$. The recursion relation for this is clearly
\beqn
{\dd\ln{g}\over \dd\ell}&=&{\dd\ln{u}\over \dd\ell}-{3\over2}{\dd\ln{c_s}\over \dd\ell}-{1\over2}{\dd\ln{c_d}\over \dd\ell}
\\
&=&{1\over u}{\dd u\over \dd\ell}-{3\over2c_s}{\dd c_s\over \dd\ell}-{1\over2c_d}{\dd c_d\over \dd\ell}
\\
&=&3-d-9g+O(g^2, mg) \,.
\label{lngrr}
\eeqn
This can trivially be rewritten 
\beqn
{\dd g\over \dd\ell}=(3-d)g-9g^2+O(g^3, mg^2) \,.
\label{grr}
\eeqn

It is convenient to choose the rescaling exponents $\chi(\ell)$ and $\zeta(\ell)$  so that at each renormalization group time $\ell$,  the parameters $c_d$ and $c_s$ remain fixed upon renormalization. This leads to two simple linear equations for 
 $\chi(\ell)$ and $\zeta(\ell)$, whose solutions are
 \beq
 \chi(\ell)=\left({1-d\over2}\right)-{g^2(\ell)\over3} \sep \zeta=2-{2g^2(\ell)\over3} \,.
 \label{csdfix}
 \eeq

Inserting these into the recursion relations \rf{mrr} and \rf{rrs fin}  for  $m(\ell)$ and $u(\ell)$ gives
\beqn
 {\dd m\over \dd\ell} &=&2m+ 6gc_s\Lambda^2-3gm+O(gm^2, g^2) \,,\nn\\
  \label{murr2_v0} 
\\
 {\dd u\over \dd\ell}  &=&(3-d-9g+O(g^2, mg))u
  \,.
 \label{murr2} 
\eeqn
 This completes our derivation of the recursion relations for the static RG.
  
 \subsubsection{The dynamic RG}{\label{drr}

If we wish to describe the {\it dynamical} (i.e., time-dependent) behavior of our system, we need to know, in addition to the renormalization of the static parameters $m$, $c_s$, $c_d$, and $u$ appearing in the Hamiltonian \rf{eq:H},  the renormalization of the kinetic coefficient $\Gamma$ appearing in the time-dependent Ginsburg-Landau equation of motion \rf{TDGL}. 

This can be done using the dynamic RG \cite{fns}. This approach is very similar to the static RG described above, with the following modifications:

\noindent 1) we must rescale time $t$ as well as the spatial coordinates and fields. We do so according to 
\beq
t\to tb^z
\label{tresc}
\eeq
where the ``dynamical" exponent $z$ is also arbitrary, and will be chosen to produce fixed points.

\noindent2) the step of averaging over the fast modes $v_x^>$ is now performed directly on the equation of motion itself.
For a detailed description of how this is done, see \cite{fns}.

This procedure has been done by \cite{folk_zpb77}, who found
\beq
{\dd\Gamma\over \dd\ell}=\left(z-\zeta-2\chi-d+1-Ag^2+O(mg^2, g^3)\right)\Gamma \,,
 \label{gammarr}
\eeq
where the constant $A$ is determined by numerically evaluating  a nasty multi-dimensional integral, and is given by
\beq
A=2.56 \,.
\label{A}
\eeq

It is also convenient to choose the arbitrary temporal rescaling exponent $z$ to keep the kinetic coefficient $\Gamma$ fixed. From
\rf{gammarr}, we see that this leads to the choice
\beq
z(\ell)=\zeta(\ell)+2\chi(\ell)+Ag^2(\ell)+d-1=2+\left(A-{4\over3}\right)g^2(\ell) \,.
\label{z}
\eeq

We'll now discuss the implications of these recursion relations for different spatial dimensions.

\subsection{Implications of the recursion relations in different spatial dimensions}

\subsubsection{$d>3$}

We can see immediately from the recursion relation \rf{grr} for the dimensionless coupling $g$ that $g(\ell)$ vanishes exponentially (like $e^{(3-d)\ell}$) for $d>3$. This implies that the effects of the non-linearity will vanish at large length scales for these higher spatial dimensions. As a result, our linear results will apply, with only finite corrections to the bare parameters of the model. In the language of the remormalization group, this implies that $d=3$ is the critical dimension for our problem.

\subsubsection{$d=3$}

The RG flows in the $g$-$u$ plane implied by the recursion relations \rf{grr}  in $d=3$ are depicted in figure \ref{d=3flows}. As can be seen from that figure, there is a single fixed point at $m=0=g$; we call this the ``Gaussian". Points   along the flow line that flows into this Gaussian fixed point form the ``critical surface". Points above this line eventually flow up to large positive $m$, and are therefore in the disordered phase, while those below the line eventually flow up to large negative $m$, and are therefore in the ordered phase. Thus, that incoming line forms a separatrix for the flows, and also is the phase boundary between the ordered and disordered phases.

By tuning experimental parameters (e.g., temperature), experimentalists (or simulators) can move the initial conditions for the RG flows along some locus, as illustrated by the dashed line in figure \ref{d=3flows}. The point at which this locus crosses the separatrix is where the phase transition occurs in the model. For values of the experimentally tuned parameter $p$  that are close to the value $p_c$ at which this crossing occurs, the experimental initial condition will lie close to the separatrix, and hence will flow very close to the Gaussian fixed point, before eventually turning away and flowing either up to large positive $m$, or down to large negative $m$, depending on which side of the separatrix the system started. One therefore expects the distance from the separatrix, for values of the experimentally tuned parameter close to $p_c$, to be proportional to $p-p_c$. We will use this fact in the analysis that follows to determine the scaling of various experimental observables with $p-p_c$.

Our analysis begins by noting that, as in $d>3$, in $d=3$ 
$g(\ell)$ still vanishes, but now much more slowly, as can be seen by solving the recursion relation \rf{grr} in $d=3$, where, for $m$ and $g$ small enough that the $O(g^3,mg^2)$ terms can be neglected, it reads
\beqn
{\dd g\over \dd\ell}=-9g^2  \,.
\label{grr3d}
\eeqn
This is readily solved:
\beq
g(\ell)={g_0\over1+9g_0\ell} \,,
\label{gsol}
\eeq
where $g_0\equiv g(\ell=0)$ is the ``bare" value of $g$; that is, its value in the original model, before we start renormalizing.

Even this slow vanishing is fast enough that the linear results for the critical point
 that we presented earlier are recovered. However, the results of the linear theory acquire logarithmically divergent corrections close to the critical point. We'll now demonstrate this.

Dropping the $O(gm^2)$ terms in \rf{murr2_v0}, that equation becomes a linear, inhomogeneous,  first order ODE in $m$. As always for such equations, the general solution is a sum of {\it any} solution to the inhomogeneous equation, plus an arbitrary solution to the homogeneous equation (which in this case means the equation obtained by dropping the $6gc_s\Lambda^2$ term). 

A simple particular solution to the inhomogeneous equation  \rf{murr2_v0} is 
\beq
m(\ell)=-3gc_s\Lambda^2+O(g^2) \,,
\label{part}
\eeq
as can be verified by direct substitution, upon noting that  \rf{part} implies
\beqn
 {\dd m\over \dd\ell} &=&-3c_s\Lambda^2 {\dd g\over \dd\ell}=27c_s\Lambda^2 g^2=O(g^2) \,,
 \nn\\
 \label{dmdlpart} 
\eeqn
where in the second equality we've used the recursion relation \rf{grr} for $g$.

Given \rf{dmdlpart}, we see that both $ {\dd m\over \dd\ell} $ and the $gm$ terms in \rf{murr2}  can be absorbed in the $O(g^2)$, while the terms linear in $g$ cancel exactly. This proves that our particular solution \rf{part} is correct.

The homogeneous ODE obtained from dropping the inhomogeneous $6gc_s\Lambda^2$ term, and the $O(g^2)$ terms, is
\beqn
 {\dd m\over \dd\ell} &=&(2-3g)m  \,.
 \label{mrr3d} 
\eeqn
This can be immediately solved by separation of variables:
\beq
m(\ell)=f(m_0)\exp\bigg[\int_0^\ell \bigg(2-3g(\ell^\prime)\bigg) \dd\ell^\prime\bigg] \,,
\label{msol3d}
\eeq
where $f(m_0)$ is a constant of integration which depends on the initial conditions of the RG flow equations, and, hence, on the ``bare" value $m_0=m(\ell=0)$ of $m$.

Inserting our solution \rf{gsol} for $g(\ell)$ into \rf{msol3d}, we see that the integral in \rf{msol3d} is elementary, and gives
\beqn
\int_0^\ell \bigg(2-3g(\ell^\prime)\bigg) d\ell^\prime&=&2\ell-3g_0\int_0^\ell{\dd\ell^\prime\over1+9g_0\ell^\prime}
\nonumber\\
&=&2\ell-{1\over3}\ln(1+9g_0\ell) \,.~~~~~~
\label{3dint}
\eeqn
Inserting this into \rf{msol3d}, and adding our particular solution \rf{part}, we obtain our general solution for $m(\ell)$:
\beq
m(\ell)=-3g(\ell)c_s\Lambda^2+{f(m_0)e^{2\ell}\over(1+9g_0\ell)^{1/3}} \,.
\label{msol3dcomp}
\eeq

Note that the solutions \rf{gsol} and \rf{msol3dcomp} are asymptotically {\it exact} for large $\ell$ 
 in $d=3$, since the higher order $O(g^2)$ terms 
that we have neglected all vanish like ${1\over\ell^2}$ at large $\ell$. Since the integral of ${1\over\ell^2}$ over $\ell$ converges as $\ell\to\infty$, those higher order terms can only contribute a finite  overall multiplicative constant to our solutions for $g(\ell)$ and $m(\ell)$. 

 \begin{figure}
    \centering
    \includegraphics[width=0.9\linewidth]{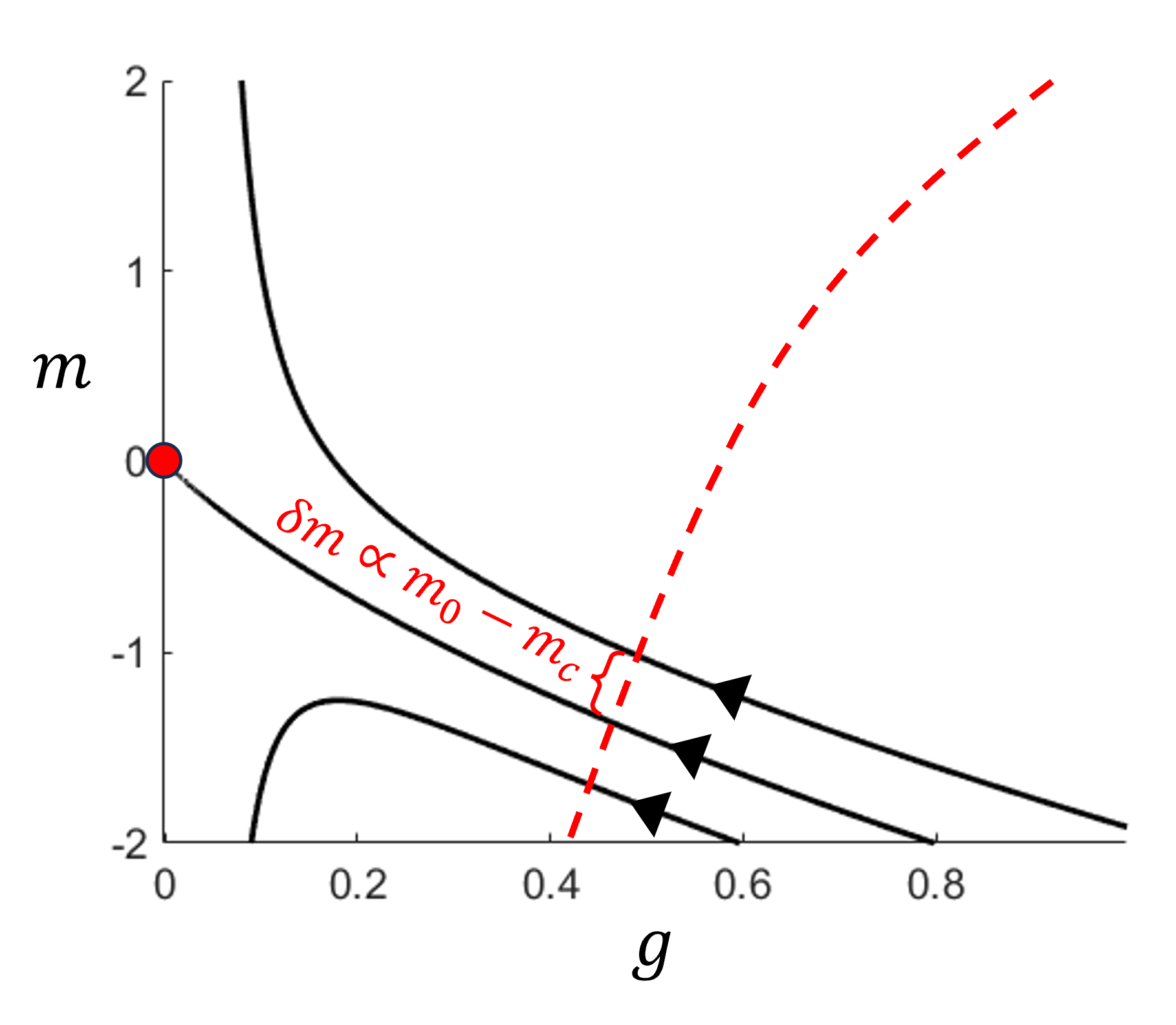}
   \caption{RG flows in the $g$-$m$ plane for spatial dimension $d=3$. The stable Gaussian fixed point is denoted by the red circle.
The flows are generated using Eqs~(\ref{grr}) \& (\ref{murr2_v0}), with the  parametrization: $3c_s\Lambda^2 =5$. 
   }
    \label{d=3flows}
\end{figure}

Since, as $\ell\to\infty$, $g(\ell)$ vanishes,  while the second term diverges exponentially, it is clear that the second term will dominate as $\ell\to\infty$, {\it unless} $f(m_0)=0$. For positive $f(m_0)$, $m(\ell)$ goes to large positive values, which, as we saw in our earlier discussion, corresponds to the ordered phase. Likewise, if $f(m_0)$ is negative, 
$m(\ell)$ goes to large negative values, which, as we saw in our earlier discussion, corresponds to the {\it dis}ordered phase.  Therefore, as we tune $m_0$,  the order-disorder transition
must occur at a value $m_c$ of $m_0$
such that $f(m_c)=0$. Since we expect $f(m_0)$ to be an analytic function of the control parameter $m_0$, this implies that, near $m_c$, 
Setting $\ell=0$ into (\ref{msol3dcomp}), we  get $m_0=m_c+f(m_0)$ with $m_c=-3g_0c_s\Lambda^2$. This implies  
\beq
\label{fcrit}
f(m_0)=(m_0-m_c)
\ .
\eeq

Recall that when $m(\ell)\to\infty$ as $\ell\to\infty$, we must be in the disordered phase, while if $m(\ell)\to-\infty$ as $\ell\to\infty$, we are in the ordered phase. Inspection of \rf{msol3dcomp} shows that this implies that $f(m_0)>0$ above the transition, while $f(m_0)<0$ below the transition. Therefore, right {\it at} the transition, we must have $f(m_0)=0$. Thus, if $m_0(p)$ is an analytic function of some experimental control parameter $p$, then $f(m_0(p))$ must also be an analytic function of $p$; furthermore, it must vanish at $p_c$, the value of the control parameter at which the transition occurs. Hence, near the transition, we must have
\beq
f(m_0(p))=(m_0-m_c)\approx C_f(p-p_c)/p_c \,,
\label{acrit}
\eeq
with $C_f/p_c\equiv {\dd f(m_0(p))\over \dd p}|_{p=p_c}$, a non-universal (i.e., system-dependent), non-zero constant.

We can also obtain the renormalized quadratic coupling
$u(\ell)$ from the solution \rf{gsol} for $g(\ell)$, using the definition \rf{gdef} of $g(\ell)$, and taking advantage of the fact that, with our convenient choice of the rescaling exponents, all of the parameters appearing in \rf{gdef} are constant under the RG, except for $g$ and $u$. Hence, we have
\beq
u(\ell)=\left({2\pi\sqrt{c_s^3c_d}\over \kbt}\right)g(\ell)={u_0\over1+9g_0\ell} \,,
\label{usol}
\eeq
where we have set $d=3$ in \rf{gdef}.

With the solution of the recursion relations \rf{gsol} and \rf{msol3dcomp}, and the values \rf{csdfix} and \rf{z} of the RG rescaling factors $\chi(\ell)$, $\zeta(\ell)$, and $z(\ell)$  in hand, we can now calculate the physical observables for the full non-linear theory.

We begin with the mean speed $|\langle v_x^{\rm Phys}(m_0)\rangle|$, which we denote with the superscript ``Phys" to distinguish it from the velocity in the renormalized system, which we'll denote by $|\langle v_x(m(\ell), u(\ell))\rangle|$. These two are related by the rescaling \rf{rescale}. Undoing the sequence of those rescalings leads to the pretentiously called ``RG trajectory integral matching formalism", which in this context implies
\beq
|\langle v_x^{\rm Phys}(m_0)\rangle|=\exp\bigg[\int_0^\ell \chi(\ell^\prime) \,\dd\ell^\prime\bigg]
|\langle v_x(m(\ell), u(\ell))\rangle| \,.
\label{trajM}
\eeq

We will now choose $\ell$ to take on a special value $\ell^*$ such that $m(\ell^*)$ as given by  (\ref{msol3dcomp}) takes on some reference value $m_{\rm ref}$ that is large. By ``large" in this context, we mean large enough that the effect of fluctuations on $\langle v_x(m(\ell^*), u(\ell^*))\rangle$ can be ignored. We will choose $|m_{\rm ref}|=c_s\Lambda^2$.

Hence, ``mean field theory" - that is, ignoring the effects of noise and the resultant fluctuations on the equation of motion \rf{V_x1} - works well for calculating $|\langle v_x(m(\ell), u(\ell))\rangle|$. This mean field theory is exceedingly simple: one simply replaces $v_x(\br,t)$ with $\langle v_x(m(\ell^*), u(\ell^*))\rangle$ in the Hamiltonian \rf{eq:H}, and then minimizes that Hamiltonian over $\langle v_x(m(\ell^*), u(\ell^*))\rangle$. The result is
\beq
|\langle v_x(m(\ell^*), u(\ell^*))\rangle|=\sqrt{-{m(\ell^*)\over u(\ell^*)}}  \,,
\label{MFTsol}
\eeq
where $m(\ell^*)<0$, since we are considering the system  in its  ordered phase.

Recall that we have chosen $\ell^*$ so that $|m(\ell^*)|=c_s\Lambda^2$,  {\it regardless of the bare value} $m_0$ of $m$. The choice of $\ell^*$ that accomplishes this obviously depends on $m_0$. From
\rf{msol3dcomp} and \rf{acrit}, we see that $\ell^*(m_0)$ must obey
\beq
|m(\ell^*)|=
{|C_f(p-p_c)/p_c|\ee^{2\ell^*}\over(\ell^*)^{1/3}}=c_s\Lambda^2 \,.
\label{ell*cond}
\eeq
This condition is easily solved for $\ell^*(p)$ in the critical limit $p\to p_c$:
\beqn
\ell^*(p)=&-&{1\over2}\ln\left({\Big|{p-p_c\over p_c}\Big|} \right)
+{1\over6}\ln\left[\ln\left({\Big|{p_c\over p-p_c}\Big|} \right)\right]
\nn\\
&+&O(1)
\,.
\label{ell*}
\eeqn
Using this value of $\ell^*$ in \rf{usol} gives
\beq
u(\ell^*)={u_0\over1+9g_0\ell^*}\approx
{2u_0\over 9g_0 \ln\left({p_c\over |p-p_c| } \right)}={2\pi\sqrt{c_s^3c_d}\over\kbt |\ln\left({p-p_c\over p_c} \right)|} \,,
\label{u*}
\eeq
where in the last equality we have used the definition \rf{gdef} of $g$ to relate $g_0$ to $u_0$.

Now we
just need to evaluate the integral in \rf{trajM}. Using our result \rf{csdfix} for $\chi(\ell)$ evaluated in $d=3$, we see that this integral, evaluated at $\ell^*$,  is
\beqn
\int_0^{\ell^*} \chi(\ell^\prime) \,\dd\ell^\prime&=&\int_0^{\ell^*}\left[-1-{g^2(\ell^\prime)\over3}\right] \,\dd\ell^\prime\nonumber\\
&=&-\ell^*-{1\over3}\int_0^{\ell^*}g^2(\ell^\prime)\,\dd\ell^\prime \,.
\label{trajint}
\eeqn
Using our expression \rf{gsol} for $g(\ell)$, we see that the integrand in the last integral vanishes like $1/\ell^{\prime\,2}$, which is fast enough that the integral coverges 
to a finite constant as $\ell^*\to\infty$.  Thus, as we approach the critical point, so that $m_0\to m_c$, and, hence, $\ell^*\to\infty$, we can asymptotically replace that last integral with its value with the upper limit taken to $\infty$, which gives
\beq
\int_0^{\ell^*}g^2(\ell^\prime)\,\dd\ell^\prime\approx \int_0^{\infty}g^2(\ell^\prime)\,\dd\ell^\prime
=\int_0^{\infty}{g_0^2\dd\ell^\prime\over (1+9g_0\ell^\prime)^2}={g_0\over9} \,.
\label{trajint2}
\eeq
 
Inserting this, \rf{trajint}, \rf{u*},  \rf{ell*cond} and \rf{MFTsol} into  \rf{trajM}  gives
 \beq
 \langle v_x^{\rm Phys} \rangle\propto |p-p_c|^{1/2} \big|\ln\left(|(p_0-p_c)/p_c|\right)\big|^{1/3} \,,
 \label{OPd=3.2}
 \eeq
which the alert reader will recognize as equation \rf{OPd=3} of the introduction. 

The scaling behavior of the two-point correlation function
\beq
C(\br,t)\equiv \langle v_x(\bR+\br,t+T)v_x({\mathbf R},T)\rangle
\label{cdef}
\eeq
 within the critical region
can also be derived using the trajectory matching method.

The correlation function
only depends on the 
relative spatio-temporal displacements $t$, $x$, and $\br_{_\perp}$, since the system is homogeneous in both
space and time. It also only depends on the {\it magnitudes} $|t|$ and $|x|$ of $t$  and $x$, since taking $t$ to $-t$ or $x$ to $-x$ is equivalent (using translation invariance in space and time) to reversing the order of  
$v_x(\bR+\br,t+T)$ and $v_x({\mathbf R},T)$ in \rf{cdef}.
Likewise, $C(\br,t)$ only depends on the magnitude 
$|\br_{_\perp}|$
of
$\br_{_\perp}$, since the system is rotation invariant about the $x$-axis.

The RG establishes, via the trajectory integral matching formalism described above, the following relation between the correlation function for the original system and that for
the rescaled system \cite{Nelson_traj}:
\bew
\begin{eqnarray}
&&C\left(\brp,x,t; m_0, u_0, \Gamma, c_s, c_d, \Lambda\right)\nonumber\\
&=&\exp\bigg[\int_0^{\ell} 2\chi(\ell^\prime)\dd\ell^\prime\bigg]
C\left[|\br_{_\perp}|e^{-\ell}, |x|\exp\bigg[-\int_0^{\ell} \zeta(\ell^\prime)\dd\ell^\prime\bigg], |t| \exp\bigg[-\int_0^{\ell} z(\ell^\prime)\dd\ell^\prime\bigg];
m(\ell), u(\ell), \Gamma, c_s, c_d, \Lambda
\right]\,,~~~~~~~
\label{Scaling_Co1}
\end{eqnarray}
\ew
where the subscripts ``0" denote the bare values of the parameters. 
 Note that  the parameters $c_s$, $c_d$, $\Gamma$, and the ultraviolet cutoff $\Lambda$ on the right-hand side of this expression are {\it not} dependent on the RG ``time``
$\ell$, since we have chosen the rescaling exponents to keep them fixed.

Now we choose to apply this expression at $\ell=\ell^*$ with $\ell^*$ given by \rf{ell*}.

With this choice of $\ell$, equation \rf{Scaling_Co1} reads
\bew
\begin{eqnarray}
&&C\left(\brp,x,t; m_0, u_0, \Gamma, c_s, c_d, \Lambda\right)\nonumber\\
&=&\exp\bigg[2\int_0^{\ell^*} \chi(\ell^\prime)\dd\ell^\prime\bigg]
C\left[|\brp| e^{-\ell^*}, |x|\exp\bigg[-\int_0^{\ell^*} \zeta(\ell^\prime)\dd\ell^\prime\bigg], |t| \exp\bigg[-\int_0^{\ell^*} z(\ell^\prime)\dd\ell^\prime\bigg];
m(\ell^*), u(\ell^*), \Gamma, c_s, c_d, \Lambda
\right]\,.~~~~~~~~~~
\label{Scaling_Co2}
\end{eqnarray}
\ew
The first integral over $\ell^\prime$ in this expression (i.e., the integral of $\chi$) is exactly the same as the integral \rf{trajint} we evaluated earlier. As noted there, this integral is given, for large $\ell^*$ (which means large $\brp$), by
\beq
\int_0^{\ell^*} \chi(\ell^\prime)\dd\ell^\prime=-\ell^*+C_\chi \,,
\label{chiintd=3.2}
\eeq
where the finite, $O(1)$, non-universal  constant 
\beq
C_\chi\equiv -{1\over3}\int_0^\infty g^2(\ell^\prime) \dd\ell^\prime
\label{cchidef}
\eeq
arises from the convergent integral of $g^2(\ell^\prime)$ up to $\ell^\prime=\infty$.

The other two integrals in our expression \rf{Scaling_Co2} similarly involve integrals of $g^2(\ell^\prime)$ up to $\ell=\infty$, which also lead to
finite additive constants. Thus we have
\beq
\int_0^{\ell^*} \zeta(\ell^\prime)\dd\ell^\prime=2\ell^*+2C_\chi\,,
\label{zetaintd=3}
\eeq
and
\beq
\int_0^{\ell^*} z(\ell^\prime)\dd\ell^\prime=2\ell^*+(4-3A)C_\chi
%= 2\ell^*-3.68C_\chi
=2\ell^*+C_z\,,
\label{zintd=3}
\eeq
where
\beq
C_z\equiv -3.68C_\chi\,.
\eeq

Inserting \rf{chiintd=3.2},
\rf{zetaintd=3}, 
and
\rf{zintd=3} into \rf{Scaling_Co2}, and using the value \rf{ell*} of $\ell^*$, we obtain 
\bew
\begin{eqnarray}
&&C\left(\brp,x,t; m_0, u_0, \Gamma, c_s, c_d, \Lambda\right)\nonumber\\
&=&e^{2C_\chi}\delta\left[|\ln\delta|\right]^{ -1/3}
C\left[r_{_\perp}\sqrt{\delta}\left[|\ln\delta|\right]^{-1/6}, |x|e^{-2C_\chi}\delta\left[|\ln\delta|\right]^{ -1/3}, |t| e^{-C_z}\delta\left[|\ln\delta|\right]^{ -1/3};
c_s\Lambda^2, u(\ell^*), \Gamma, c_s, c_d, \Lambda
\right]\,.~~~~~~~~~~~
\label{Scaling_Co3}
\end{eqnarray}
\ew
where we've defined
\beq
\delta\equiv {|m_0-m_c|\over c_s\Lambda^2}\approx{1\over c_s\Lambda^2}\Bigg|{C_f(p-p_c)\over p_c}\Bigg|
\label{deltadef}
\eeq
which is a natural measure of the system's proximity to the transition. In the approximate equality in \rf{deltadef}, we have used the relations between the parameters given in \rf{acrit}. In writing \rf{Scaling_Co3}, we have assumed that the system is in the disordered phase so that $ m(\ell^*)= c_s\Lambda^2$, otherwise $ m(\ell^*)= -c_s\Lambda^2$ if the system is in the ordered phase.

To proceed further, we note that, when we are close to the transition - that is, when $\delta$ is small - the anharmonic coefficient $u(\ell^*)$ is also small, as can be seen from \rf{u*}, while the mass  $|m(\ell^*)|$ is $c_d$,  which, in appropriate units, is  $O(1)$.  The latter fact implies that the fluctuations of $v_x$ will not be particularly large. Taken together, these facts imply that the quartic $u$ term in the Hamiltonian \rf{eq:H} can be neglected. This in turn implies that the  correlation function on the right hand side of \rf{Scaling_Co3} can be evaluated using the linear theory  presented in section \rf{lin} for systems in the disordered phase. 

Doing this initially for the equal-time correlation function in the disordered phase, we have 
\bew
\begin{eqnarray}
&&C\left[r_{_\perp}\sqrt{\delta}\left[|\ln\delta|\right]^{-1/6}, |x|e^{-2C_\chi}\delta\left[|\ln\delta|\right]^{-1/3}, t=0;
c_d, u(\ell^*), \Gamma, c_s, c_d, \Lambda
\right]
\nn\\
&=&
{D_x\xpe^{\rm ref}\left(2{a(\ell^*)\over w}x^2\ee^{-4C_\chi}\delta^2\left[|\ln\delta|
\right]^{-2/3}-r_{_\perp}^2\delta\left[|\ln\delta|\right]^{-1/3}\right)\over 4\pi\sqrt{\mup w}
\bigg(r_{_\perp}^2\delta\left[|\ln\delta|\right]^{-1/3}+{a(\ell^*)\over w}x^2\ee^{-4C_\chi}
\delta^2\left[|\ln\delta|\right]^{-2/3}\bigg)^{5/2}}\nn\\
&=& {D_x\sqrt{|
\ln{\delta}|}\left(2x^2\ee^{-4C_\chi}
\delta\left[|\ln\delta|\right]^{-1/3}-r_{_\perp}^2\right)\over 4\pi\sqrt{\mu_\perp w \delta^3}\Lambda\bigg(r_{_\perp}^2+x^2\left({\mu_\perp\over w}\Lambda^2\right)\ee^{-4C_\chi}\delta\left[|
\ln\delta|\right]^{-1/3}\bigg)^{5/2}}
\,,~~~~~~~~~
\label{Scaling_Cet}
\end{eqnarray}
\ew
where we have used  the relations \rf{eqcon} between the Hamiltonian parameters $m$ and $c_d$ and the dynamical parameters $a$ and $w$ appearing in the expressions
\rf{eq:vx4} and \rf{f2d}  for the equal-time correlation function to write ${a(\ell^*)\over w}={m(\ell^*)\over c_d}={c_s\Lambda^2\over c_d}={\mu_\perp\over w}\Lambda^2$, where the last two equalities follow from choosing 
$\ell^*$ 
so that $m(\ell^*)=c_s\Lambda^2$. Similar logic tells us that 
\beq
\xpe^{\rm ref}=\sqrt{\mu_{_\perp}\over a(\ell^*)}=\sqrt{\Gamma c_s\over\Gamma m(\ell^*)}=\sqrt{c_s\over c_s\Lambda^2}
={1\over\Lambda} \,,
\label{xiref}
\eeq
where in the second equality we have used $a(\ell^*)=w$, which follows from $a(\ell^*)=\Gamma m(\ell^*)=\Gamma c_s\Lambda^2$.
This is a microscopic length, which is non-singular as the transition is approached; that is, as $m_0\to m_c$.

Inserting \rf{Scaling_Cet} into \rf{Scaling_Co3} gives 
\begin{eqnarray}
&&C\left(\brp,x,t=0; m_0, u_0, \Gamma, c_s, c_d, \Lambda\right)\nonumber\\
&=&{D_x\xpe(m_0(p))\left(2\alpha x^2
-r_{_\perp}^2\right)\over 4\pi \sqrt{\mu_{_\perp}w}(r_{_\perp}^2+\alpha x^2)^{5/2}}
\,,
\label{Scaling_Cetfin}
\end{eqnarray}
where we've
defined the perpendicular correlation length 
\beqn
\xpe(m_0(p))
&\equiv&{
\ee^{2C_\chi}|\ln{\delta}|^{1/6}\over\Lambda\sqrt{\delta} }\nonumber\\
&\approx&{
\ee^{2C_\chi}\Big|\ln\left(\big|{C_f(p-p_c)\over c_s\Lambda^2p_c}\big|\right)\Big|^{1/6}\over\Lambda\sqrt{\Big|{C_f(p-p_c)\over c_s\Lambda^2p_c}\Big|} }\nonumber\\
&\propto&|p-p_c|^{-{1\over 2}}
\bigg|\ln\left(\Big|{p-p_c\over p_c}\Big|\right)\bigg|^{1/6}
\label{xpedef}
\eeqn
and 
\beq
\alpha=\left({\mu_\perp\over w}\Lambda^2\right)\ee^{-4C_\chi}
\delta|\ln\delta|^{-1/3}
={\mu_{_\perp}\over w\xpe^2(p)} \,,
\label{alphadef}
\eeq
where in the ``$\approx$" we have used \rf{deltadef}.

Since we have chosen our RG rescaling factors to keep as many of the parameters fixed as possible, the only singular dependence of this correlation function near the transition is in the explicitly displayed $a_0$ dependence of $\xpe(a_0)$.

The alert reader will recognize these expressions \rf{Scaling_Cetfin} and \rf{xpedef} as the results \rf{xilog_perp}, \rf{eq:vx3I}, and \rf{f3dI} for the correlation $\xpe$ length and  equal-time correlation function in the non-critical regime quoted in the introduction.

The above analysis only applies if {\it at least one} of the conditions  $|\brp|\gg\xpe(m_0(p))$, $|x|\gg\xpa(m_0(p))$ is satisfied, 
where $\xpa$ is defined as the characteristic length of $x$ such that the rescaled length $ |x|e^{-2C_\chi}\delta|\ln\delta|^{-1/3}$ is of order $O(\sqrt{w/\mu_{_\perp}}/\Lambda^2)$. This definition of $\xpa$ leads to 
\beqn
\xpa(m_0(p))&\equiv&{{\sqrt{\mu_{_\perp}\over w}}\ee^{2C_\chi}|\ln{\delta}|^{1/3}\over\delta\Lambda^2 }
\nn\\
&\approx&
{\sqrt{\mu_{_\perp}\over w}
\ee^{2C_\chi}\Big|\ln\left(\big|{C_f(p-p_c)\over c_s\Lambda^2p_c}\big|\right)\Big|^{1/3}\over\Big|{C_f(p-p_c)\over c_sp_c}\Big| }
\nonumber\\
&\propto&
|p-p_c|^{-1}
\bigg|\ln\left(\Big|{p-p_c\over p_c}\Big|\right)\bigg|^{1/3}
\label{xpadef}
\eeqn
which is essentially the result \rf{xilog_para} given in the introduction.
The reason for these limitations on $\brp$ and $x$ is that we have implicitly assumed in the above argument that at least one of the rescaled lengths $r_{_\perp}\sqrt{\delta}\left[|\ln\delta|\right]^{-1/6}$ and $ |x|e^{-2C_\chi}\delta\left[|\ln\delta|\right]^{-1/3}$ is large enough that our hydrodynamic theory, which is only valid on long length scales, can be used to calculate the rescaled correlation function on the right hand side of \rf{Scaling_Co2}. 

If this is not the case, then we have to proceed by making a choice  different from from \rf{ell*} of the value of $\ell$ at which to apply \rf{Scaling_Co2} . We'll discuss this choice in more detail below.

We now turn to the 
equal-space, time-dependent correlation function 
$C\left(\brp={\bf 0},x=0,t; m_0, u_0, \Gamma, c_s, c_d, \Lambda\right)$ in the non-critical regime. From the general scaling relation \rf{Scaling_Co3}, we see that this is given by
\bew
\begin{eqnarray}
C\left(\brp={\bf 0},x=0,t; m_0, u_0, \Gamma, c_s, c_d, \Lambda\right)
&=&\ee^{2C_\chi}\delta\left[|\ln\delta|\right]^{-1/3}
C\left[0, 0, |t| \ee^{-C_z}\delta\left[|\ln\delta|\right]^{-1/3};
c_s\Lambda^2, u(\ell^*), \Gamma, c_s, c_d, \Lambda
\right]\,.\nn\\
\label{Scaling_C(t)}
\end{eqnarray}
\ew
Once again, near the transition, where $\ell^*$ is large, we can evaluate the correlation function the the right hand side using the linear theory of section \rf{lin}, specifically, the $|t|\gg
\tau_{\rm corr}$
limit  (that is, the non-critical limit) of \rf{Cet3dnoncrit}. This gives
\begin{eqnarray}
\nn
&&C\left(\brp={\bf 0},x=0,t; m_0, u_0, \Gamma, c_s, c_d, \Lambda\right)\\
&=&{D_x\tau_{\rm corr}(m_0(p))B_t\over16\pi\sqrt{w\mu_\perp^3}}{\ee^{-|t|/\tau_{\rm corr}}\over t^2}
\label{Cetsc}
\eeqn
where we've defined the correlation time 
\beqn
\tau_{\rm corr}(m_0(p))&\equiv&{|\ln\delta|^{1/3}\over\Gamma c_d\delta }\ee^{C_z}
\nn
\\
&\approx&{\ee^{C_z}\bigg|\ln\left(\big|{C_f(p-p_c)\over c_dp_c}\big|\right)\bigg|^{1/3}\over \Gamma\big|{C_f(p-p_c)\over p_c}\big|}
\nn\\
&\propto&|p-p_c|^{-1}{\bigg|\ln\left(\Big|{p-p_c\over p_c}\Big|\right)\bigg|^{1/3}}
\label{tauc}
\eeqn
and the $O(1)$ constant
\beq
B_t\equiv \ee^{2C_\chi+C_z} \,.
\label{btdef}
\eeq
Equation \rf{Cetsc} is, of course, just our result \rf{Cet3I} quoted in the introduction, and \rf{tauc} confirms our prediction in equation \rf{xilog_para} of the introduction for the critical behavior of the correlation time. 

As we argued earlier for the length scales, so here this non-critical calculation only applies for $|t|\gg\tau_{\rm corr}$; otherwise, time argument of  the correlation function on the right hand side of \rf{Scaling_C(t)}  is too small for us to use our hydrodynamic theory. So in this case too, we'll have to make a different choice of the value of $\ell$ at which to apply \rf{Scaling_Co1}. 

We'll now describe that choice, starting with the equal-time correlation function.

If $|\brp|\ll\xi_\perp$, we choose
\beq
\ell=\ell_s(r_{_\perp})=\ln\left(\Lambda r_{_\perp}\right) \,.
\label{lr}
\eeq 
Note that $\ell_s(r_{_\perp})$ will be large if $|\brp|\gg\Lambda^{-1}$, but at the same time, we`ll have $\ell_s(r_{_\perp})\ll\ell^*$ if $r_{_\perp}\ll\xpe$.

With this choice of $\ell_s(r_{_\perp})$ in \rf{Scaling_Co1}, we have
\bew
\begin{eqnarray}
&&C\left(\brp,x,t; m_0, u_0, \Gamma, c_s, c_d, \Lambda\right)
\nonumber\\
&=&e^{2C_\chi}(\Lambda r_{_\perp})^{-2}
C\left[\Lambda^{-1}, |x|e^{-2C_\chi}(\Lambda r_{_\perp})^{-2}, |t| e^{-C_z}(\Lambda r_{_\perp})^{-2};
m(\ell_s(r_{_\perp})), u(\ell_s(r_{_\perp})), \Gamma, c_s, c_d, \Lambda
\right]\,.
\label{Scaling_Cocrit}
\end{eqnarray}
\ew

As before, when $\ell_s(r_{_\perp})$ is large (specifically, $\ell_s(r_{_\perp})\gg1$), $u(\ell_s(r_{_\perp}))$ is small. At the same time, since   
$\ell_s(r_{_\perp})\ll\ell^*$ for $r_{_\perp}\ll\xpe$, then for this range  
$\Lambda^{-1}\ll\rp\ll\xpe$, we`ll have $|m(\ls)|\ll c_s\Lambda^2$, as can be seen from the fact that $m(\ell)$ is (at large $\ell$) a monotonically increasing function of $\ell$, and is equal to $c_s\Lambda^2$ at $\ell=\ell^*\gg\ls$.

Because both $|m(\ls)|$ and $u(\ls)$ are small, we can evaluate the correlation function on the right hand side of \rf{Scaling_Cocrit} using the linear theory in the critical regime presented in section \rf{lin}. It is straightforward to verify that doing so for $\br={\bf 0}$  recovers the $|t|\ll\tau_{\rm corr}$ limit of our result  \rf{Cet3I} of the introduction, while setting $t=0$ we recover \rf{rscorrd=3_I}.

\subsubsection{$d=3-\epsilon$}

 \begin{figure}
    \centering
    \includegraphics[width=0.9\linewidth]{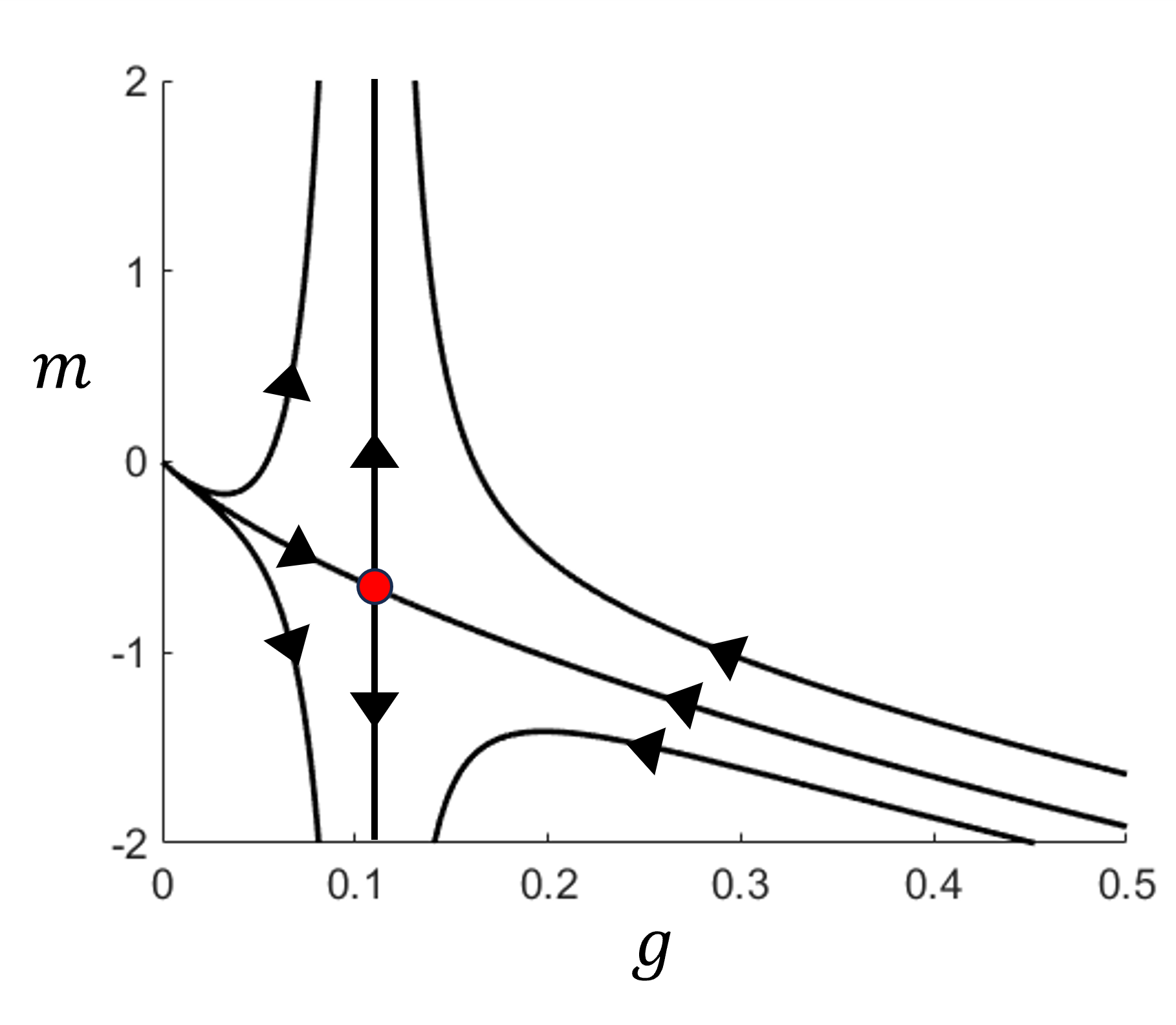}
   \caption{RG flows in the $g$-$m$ plane for spatial dimension $d=2$. The stable non-Gaussian fixed point is denoted by the red circle.
 The flows are generated using Eqs~(\ref{grr}) \& (\ref{murr2_v0}), with the terms in $\cO(.)$ omitted and choosing $3c_s\Lambda^2 =5$. 
   }
    \label{d<3flows}
\end{figure}

For $d<3$, the Gaussian $g=0$ fixed point that controls the transition for $d\ge3$  becomes unstable in both directions, with eigenvalues $3-d$ and $2$ , as can be seen immediately from equations \rf{grr} and \rf{murr2_v0}. The resultant RG flows now look like figure 
\rf{d<3flows}, with a new  fixed point  unstable in only one direction, appearing at
\beqn
\begin{aligned}
g&=g^*={\epsilon\over9}+O(\epsilon^2) \\ 
m&=m^*=-3g^*c_s\Lambda^2=-{\epsilon c_s\Lambda^2\over3}+O(\epsilon^2) \,, 
\end{aligned}
\label{3-dfp}
\eeqn
where we've defined $\epsilon\equiv3-d$, and we will follow the familiar critical phenomenon approach of treating $\epsilon$ as a small quantity, and working to leading order in it.

This structure of the renormalization group flows is the same as that of the conventional Wilson-Fisher fixed point \cite{KWF}, although the precise values of the exponents are obviously different. As in the $d=3$ case, the flow lines leading into the new, non-Gaussian fixed point \rf{3-dfp} form a separatrix between flows that ultimately go to large positive $m$, which implies the system is in the disordered phase, and flows going to large, negative $m$, which implies that the system is in the ordered phase. 
In contrast to the $d=3$ case, however, the flows near the new fixed point are exponential in RG time $\ell$, rather than algebraic. This implies that the logarithmic corrections we found in the 3D problem are replaced by power laws of the type found in conventional critical phenomena.

To show this in detail, we begin by linearizing the recursion relations \rf{murr2_v0} and \rf{grr} about the fixed point \rf{3-dfp}. We do so by writing
\beq
m(\ell)=m^*+\delta m(\ell) \sep g(\ell)=g^*+\delta g(\ell)
\label{lin_1}
\eeq
and expanding the recursion relations  \rf{murr2_v0} and \rf{grr}  to linear order in the departures $\delta m$ and $\delta g$. We find
\beqn
{\dd\dm\over \dd\ell}&=&(2-{\epsilon\over3})\delta m+6c_s\Lambda^2 \dg+O(\epsilon^2)\dm+O(\epsilon)\dg \,,\nn\\
\label{rrlin_m}
\\
{\dd\dg\over \dd\ell}&=&-\epsilon\dg+O(\epsilon^2)\dm+O(\epsilon^2)\dg \,.
\label{rrlin}
\eeqn
Seeking solutions to this linearized system of an exponential form; i.e., 
\beqn
            \delta m(\ell) =S_m\ee^{\kappa\ell} \sep g(\ell)=S_g\ee^{\kappa\ell}  \,,
\label{ev1}
\eeqn
where ${\bf S}=(S_m, S_g)$ is a constant eigenvector, and $\kappa$ a constant growth rate,  we see that there are two eigenvalues for $\kappa$:
\beqn
    \kappa_1 &= &-\epsilon+O(\epsilon^2)\\
    \kappa_2 &\equiv&\kappa_t= 2 - \frac{1}{3}\epsilon+O(\epsilon^2)
    \label{rgevs}
\eeqn
We identify $\kappa_t$, which is the only positive eigenvalue,  as the ``thermal" eigenvalue. We use this term in the usual RG sense, which is that it determines the dependence of the correlation length on the departure of the control parameter (which is usually temperature in equilibrium problems, hence the term ``thermal eigenvalue") from its critical value.

We can  see this by the following completely standard RG analysis: 

Note that the general solution of our linearized recursion relations for $\dm(\ell)$ is
\beqn
\dm(\ell)    =
    a(m_0)  e^{-\epsilon\ell}+ f(m_0)  e^{\kappa_t\ell} \,,
    \label{linsolgen}
\eeqn
where  the constants 
   $a(m_0)$ and $f(m_0)$ are determined by the initial (i.e., bare) parameters of the model, and hence, in particular, the bare value $m_0$ of $m$.

As in our discussion of the 3D case, $f(m_0)$ vanishes right at the critical point (i.e.,when the experimental control parameter $p$ reaches the critical value $p_c$). We expect that near the critical point 
\beq
f(m_0(p))\equiv m_0-m_c\approx C_f(p-p_c)/p_c \,,
\label{fcrit1}
\eeq
where $C_f/p_c\equiv {\dd f(m_0(p))\over \dd p}|_{p=p_c}$ is a non-universal positive constant.

Recognizing this, we can now use the expansion \rf{fcrit1} 
to obtain the behavior of the correlation length near the critical point by the following very standard RG argument, which is similar to, but actually simpler than, the argument we used in the 3D case.

Starting with any $m_0$ near $m_c$, we run the renormalization group until we reach a value $\ell^*(m_0)$ of $\ell$ at which the renormalized $\delta m(\ell^*)$ takes on some particular 
reference value which we'll call $\delta m_r$, which is
 small enough that the linearized recursion relations   (\ref{rrlin_m}) and (\ref{rrlin}), and, therefore, their solution  (\ref{linsolgen}), remain valid, but as big as it can be consistent with that requirement. 

For $m_0$ close to $m_c$, therefore, the value of $\ell^*$ required to reach such a large $\delta m(\ell^*)$ will clearly be large, since the coefficient $f(m_0)$ of the exponentially growing part of the solution  (\ref{linsolgen}) of our linearized recursion relations is small in that case. It follows that, by the time $\ell$ reaches $\ell^*$, the exponentially decaying $a$ term in the solution  (\ref{linsolgen}) will have become negligible, so that 
\beq
\delta m(\ell^*)\approx f(m_0)e^{\kappa_t\ell^*}=\left(\delta m\right)_r \,,
\label{ell* cond}
\eeq
where in the last equality we have applied our condition on $\ell^*$ that it make $\delta m(\ell^*)=(\delta m)_r$.
It is clearly straightforward to solve \rf{ell* cond} for $\ell^*$; we'll instead solve it for $e^{\ell^*}$, which, as we'll see in a moment, proves to be the more useful quantity:
\beq
e^{\ell^*}=\left({\left(\delta m\right)_r\over f(m_0)}\right)^{\nu_{_\perp}}\propto |m_0-m_c|^{-\nu_{_\perp}}
\propto |p-p_c|^{-\nu_{_\perp}}\,,
\label{ell*sol}
\eeq
where in the second proportionality we have used \rf{fcrit1}.  The ``correlation length exponent" is defined as
\beqn
    \nu_\perp &\equiv \frac{1}{\kappa_2}= \frac{1}{2} + \frac{\epsilon}{12} + O(\epsilon^2)  \,.  
    \label{nuperpeps} 
\eeqn
Note that, in the limit $m_0\to m_c$, all of our starting systems have been mapped onto the same point
\beq
m=m(\ell^*)=m^*+\left(\dm\right)_r \sep g=g^*
\label{refpt}
\eeq
since the exponentially decaying part of the solution  (\ref{linsolgen}) will have vanished in this limit.
Hence, all of these systems are mapped onto the same model, and, hence, onto a model with the same correlation lengths in the perpendicular and parallel direction. Those correlation lengths, which we'll call $\xpe(m(\ell^*))$ and $\xpa(m(\ell^*))$ are therefore also independent of $m_0$. 

Note that this does {\it not} imply that all of these systems have the same correlation lengths. On the contrary, since each of them will have to have been renormalized for a different, strongly $|p-p_c|$-dependent RG time $\ell^*$, as implied by  (\ref{ell*sol}), they will have very different correlation lengths. Indeed, since, on every time step, we rescale lengths in the $\perp$-directions by a factor of $b=1+d\ell$, while directions in the $\parallel$-direction are rescaled (at the one loop order to which we've worked here) by a factor of $b^{\zeta(\ell)}$, the actual correlation lengths in the $\perp$  and  $\parallel$ directions $\xi_{\parallel,\perp}$ are related to those at the ``reference point ``
by
\beqn
\xi_{\perp}(m_0(p))&=&b^{n^*}\xi_{\perp}( m(\ell^*))=e^{\ell^*}\xi_{\perp}(m(\ell^*))
\nonumber\\
&\propto&|m_0-m_c|^{-\nu_{_\perp}}\propto|p-p_c|^{-\nu_{_\perp}} \,, ~~~~~
\label{xispar}
\\
\xi_{\parallel}(m_0(p))&=&\prod_{i=1}^{n}b^{\zeta(n)}\xi_{\parallel}(m(\ell^*))\nonumber\\
&=&\exp\bigg[\int_0^{\ell^*}\zeta(\ell)\,d\ell\bigg]\xi_{\parallel}(m(\ell^*))\,.
\label{xis}
\eeqn
Here $n^*=\ell^*/d\ell$ is the number of RG steps required to reach $\ell^*$, and we've used the fact that, for $d\ell$ differential and $\ell^*=nd\ell$ finite, $b^{n^*}=(1+d\ell^*)^{n^*}
=(e^{d\ell^*})^{n^*}=e^{n^*d\ell^*}=e^{\ell^*}$. 

To proceed with the expression \rf{xis} for the parallel correlation length $\xi_\parallel$, we use the expression \rf{csdfix} for the rescaling exponents $\zeta$. Since $g$ now flows to a fixed value $g^*$ near the transition, this exponent becomes a constant  given by
\beq
\zeta =2-{2(g^*)^2\over3}=2-{\eta\over2} \,,
 \label{zetaeps}
 \eeq
where we`ve defined
\beq
\eta \equiv{4(g^*)^2\over3} =\frac{4}{243}\epsilon^2\ .
\eeq
Using this constant value of $\zeta$ in \rf{xis} gives
\beq
\xi_{\parallel}(m_0(p))\propto|m_0-m_c|^{-\zeta\nu_{_\perp}}\propto|p-p_c|^{-\zeta\nu_{_\perp}}\nn\\\equiv|p-p_c|^{-\nu_\parallel}
\eeq
with
\beq
\nu_{\parallel}\equiv\zeta\nu_{_\perp} \,,
\label{nupardef}
\eeq
which is exact to all orders in $\epsilon$.

The universal critical exponents $\nu_{\perp,\parallel}$ are given by
\beqn
    \begin{aligned}
    \nu_\perp &= \frac{1}{\kappa_2}= \frac{1}{2} + \frac{\epsilon}{12} + O(\epsilon^2) \\  \nu_\parallel &= 1 + \frac{\epsilon}{6} + O(\epsilon^2)  \,.  
    \end{aligned}
    \label{nuperpeps} 
\eeqn

By computing the higher order $\epsilon^2$ corrections to the model, Brezin et.~al.~\cite{brezin_prb76} found the values of $\nu_\perp$ and $\nu_\parallel$ to $O(\epsilon^2)$ given by equations \rf{eq:nupe} and \rf{eq:nupa} of the introduction.

Near the critical point in the ordered state the average value of $v_x$ as a function of the deviation from the critical point (i.e., $|p-p_c|$) can be obtained as we did in $d=3$; the only difference is that now the scaling exponent $\chi$ obtained from equation \rf{csdfix} is a constant given by
\beq
\chi={1-d\over2}-{(g^*)^2\over3}=-1+{\epsilon\over2}-{\eta\over4}=-1+{\epsilon\over2}- \frac{\epsilon^2}{243} +O(\epsilon^3)  \,.
\label{chieps}
\eeq

In addition, for large $\ell^*$, we will now have 
\beqn
u(\ell^*)=u^*\equiv \left({2\pi\sqrt{c_s^3c_d}\over \kbt}\right)g^* \,;
\label{u*eps}
\eeqn
that is, both $m(\ell^*)$ and $u(\ell^*)$ will go to constants, independent of how close we are to the transition. Hence,
\beq
|\langle v_x(m(\ell^*), u(\ell^*))\rangle|=\sqrt{-{m(\ell^*)\over u(\ell^*)}}
\,.
\label{MFTsol1}
\eeq
also goes to a finite constant, independent of our proximity to the transition.

Using this constant value \rf{chieps} for $\chi$  and this result \rf{MFTsol} in the trajectory integral matching expression \rf{trajM}, we find
\begin{eqnarray}
|\langle v_x^{\rm Phys}(m_0(p))\rangle|&=&e^{\chi\ell^*}|\langle v_x(m(\ell^*), u(\ell^*))\rangle|
\nonumber\\
          &\propto&|(p-p_c)|^{-\nu_{_\perp}\chi} %{\rm constant}
         \nonumber\\
          &\equiv&|p-p_c|^{\beta}\,,
\end{eqnarray}
where
\begin{eqnarray}
\beta=-\nu_{_\perp}\chi\
=\nu_{_\perp}\left({d-1\over 2}+{\eta\over 4}\right)\,.\label{beta}
\end{eqnarray}
Using our epsilon expansion values \rf{chieps} and \rf{eq:nupe} for the exponents $\chi$ and 
$\nu_{_\perp}$
gives the $\epsilon$-expansion result \rf{betaintro} for $\beta$ quoted in the introduction.

We see that the dynamical exponent $z$ given by equation \rf{z} also goes to a constant in $d=3-\epsilon$ dimensions, which is given by
\beq
z=2+\left(A-{4\over3}\right)(g^*)^2=2+\left(A-{4\over3}\right)\left({3\eta\over4}\right)=2+c\eta
\label{zeps}
\eeq
where  $c=({3A\over4}-1)\approx 0.92$ \cite{folk_zpb77}.

We can also analyse 
the two point correlation function $C\left(\brp,x,t; m_0, u_0, \Gamma, c_s, c_d, \Lambda\right)$
as we did in the 3D case, but now taking advantage of the simplifications coming from the fact that the scaling exponents $z$, $\zeta$, and $\chi$ are all constants. Using this and the value $\ell^*$ \rf{ell*sol} for $\ell$ in \rf{Scaling_Co2} gives 
\begin{widetext}
\begin{eqnarray}
C\left(\brp,x,t; m_0, u_0, \Gamma, c_s, c_d, \Lambda\right) 
=\ee^{2 \chi\ell^*}
C\left[|\brp| \ee^{-\ell^*}, |x|\ee^{-\ \zeta\ell^*}, |t| \ee^{-z\ell^*};
m(\ell^*), u^*, \Gamma, c_s, c_d, \Lambda
\right]\,.
\label{Scaling_Coeps}
\end{eqnarray}
\end{widetext}
Since $m(\ell^*)$ is $O(1)$, we can evaluate the correlation function on the right-hand side in $d=2$ using the linearized results \rf{eq:vx4} and \rf{f2d}, which taken together imply for the equal-time correlation function
\begin{widetext}
\beqn
C\left[|\brp| \ee^{-\ell^*}, |x|\ee^{-\ \zeta\ell^*}, t=0;
m(\ell^*), u^*, \Gamma, c_s, c_d, \Lambda
\right]
=
{
\xpe(m(\ell^*)) D_x\over 2\pi\sqrt{\mu_{_\perp} w}
}
{(\alpha(m(\ell^*))x^2\ee^{-2\zeta\ell^*}-r_{_\perp}^2\ee^{-2\ell^*})
\over(r_{_\perp}^2\ee^{-2\ell^*}+\alpha(m(\ell^*))x^2\ee^{-2\zeta\ell^*})^2}
\,.
\label{cscd=2}
\eeqn
\end{widetext}
Inserting this into \rf{Scaling_Coeps} gives, after a bit of algebra,
\begin{widetext}
\begin{eqnarray}
C\left(\brp,x,t=0; m_0, u_0, \Gamma, c_s, c_d, \Lambda\right)
=
\ee^{2 (1+\chi)\ell^*}
{
\xpe(m(\ell^*)) D_x\over 2\pi\sqrt{\mu_{_\perp} w}
}
{(\alpha(m(\ell^*))x^2\ee^{2(1-\zeta)\ell^*}-r_{_\perp}^2)
\over(r_{_\perp}^2+\alpha(m(\ell^*))x^2\ee^{2(1-\zeta)\ell^*})^2}
\,.
\label{Scaling_Coeps2}
\end{eqnarray}
\end{widetext}

In $d=2$, or, equivalently, $\epsilon=1$,  our result \rf{chieps} for $\chi$ implies that $1+\chi={1\over2}-{\eta\over4}$, while our result \rf{zetaeps} for $\zeta$ implies that $1-\zeta=-1+{\eta\over2}$. Using these results in \rf{Scaling_Coeps2} gives, after a little more reorganizing,
\begin{eqnarray}
\nn
&&C\left(\brp,x,t=0; m_0, u_0, \Gamma, c_s, c_d, \Lambda\right)\\
&=&
{
\xpe(m_0(p)) D_x\over 2\pi\sqrt{\mu_{_\perp}(m_0(p)) w}
}
{(\alpha(m_0(p))x^2\ee^{2(1-\zeta)\ell^*}-r_{_\perp}^2)
\over(r_{_\perp}^2+\alpha(m_0(p))x^2\ee^{2(1-\zeta)\ell^*})^2}
\,.~~~~~~\nn\\
\label{Scaling_Coeps3}
\end{eqnarray}
where the renormalized perpendicular correlation length
$\xpe(m_0(p))$ is given by equation \rf{xispar}, while the renormalized $\alpha(m_0(p))$ and $\mu_{_\perp}(m_0(p))$ are given by
\beqn
\alpha(m_0(p))&=&\alpha(m(\ell^*))\ee^{(\eta-2)\ell^*}\propto \xpe^{\eta-2}\propto|p-p_c|^{(2-\eta)\nu_{_\perp}} \,,
\nn\\
\label{aplharen}
\\
\mu_{_\perp}(m_0(p))&=&\ee^{\eta\ell^*}\mu_{_\perp}\propto \xpe^\eta\propto |p-p_c|^{-\eta\nu_{_\perp}} \,.
\label{mualpha}
\eeqn
Note that after the first equality for $\mu_{_\perp}(m_0(p))$, the quantity $\mu_{_\perp}$ is the {\it bare} value of $\mu_{_\perp}$, which we kept fixed upon renormalization, and which, therefore, has no singular behavior at the transition; all of the singular behavior of the renormalized $\mu_{_\perp}(m_0(p))$ is contained in the $ \xpe^\eta$ factor explicitly displayed.

This
recovers the results \rf{eq:vx4I} and \rf{f2dI} of the introduction.

Now we turn to the equal-position correlation. Setting $\br={\mathbf 0}$ and evaluating the correlation function on the right-hand side of \rf{Scaling_Coeps} using the result \rf{Cet2dnoncrit} in the linearized theory, we obtain 
\begin{widetext}
\beq
C\left[\br_{_\perp}={\mathbf 0}, x=0, |t|\ee^{-z\ell^*};
m(\ell^*), u^*, \Gamma, c_s, c_d, \Lambda
\right]=
{D_x\over 4\sqrt{\pi^3\mu_{_\perp}w}\xi_{_\perp}(m(\ell^*))}
\left[|t|\ee^{-z\ell^*}\over\tau_{\rm corr}(m(\ell^*))\right]^{-{3\over 2}}
\exp\left[-|t|\ee^{-z\ell^*}\over \tau_{\rm corr}(m(\ell^*))\right]\,.
\label{Scaling_Coept2}
\eeq
\end{widetext}

Inserting \rf{Scaling_Coept2} into \rf{Scaling_Coeps} gives, after a little algebra: 
\begin{widetext}
\beq
C\left(\brp={\mathbf 0},x=0,t; m_0, u_0, \Gamma, c_s, c_d, \Lambda\right) 
={D_x\over 4\sqrt{\pi^3\mu_{_\perp}(m_0(p))w}\xi_{_\perp}(m_0(p))}
\left[|t|\over\tau_{\rm corr}(m_0(p))\right]^{-{3\over 2}}
\exp\left[-|t|\over \tau_{\rm corr}(m_0(p))\right]
\label{Scaling_Coept3}
\eeq
\end{widetext}
where $\xi_{_\perp}(m_0(p))$ and $\tau_{\rm corr}(m_0(p))$ are respectively given by \rf{xispar} and \rf{mualpha}, and
\beq
\tau_{\rm corr}(m_0(p))\equiv \tau_{\rm corr}(m(\ell^*)\ee^{z\ell^*}
\propto\xi_{_\perp}^z
\propto|p-p_c|^{-z\nu_{_\perp}}\,.
\eeq
In deriving \rf{Scaling_Coept3} we have used $2\chi=-1-{\eta\over 2}$ in $d=2$. We recognize that these results are just \rf{eq:vx5I} and \rf{taudiv} in the introduction.

As in $d=3$, for $|\brp|\ll\xi_\perp$, we make  the  choice \rf{lr} of the value of $\ell$ at which to apply equation \rf{Scaling_Co1}.
Doing so, we`ll again have $1\ll\ls\ll\ell^*$. In this case, this means that $m(\ls)$ and $u(\ls)$ will have flowed to their fixed point values $m^*$ and $u^*$ given by eqns. \rf{3-dfp} and \rf{u*eps}.
Hence we have 
\bew
\begin{eqnarray}
C\left(\brp,x,t; m_0, u_0, \Gamma, c_s, c_d, \Lambda\right)
%\nonumber\\
=(\Lambda r_{_\perp})^{2\chi}
C\left[\Lambda^{-1}, |x|(\Lambda r_{_\perp})^{-\zeta}, |t| (\Lambda r_{_\perp})^{-z};
m^*, u^*, \Gamma, c_s, c_d, \Lambda
\right]\,,
\nn\\
\label{Scaling_Cocriteps}
\end{eqnarray}
\ew
with $\chi$, $\zeta$, and $z$ now being the constant universal exponents given by eqns. \rf{chieps}, \rf{zetaeps}, and \rf{zeps}.

Note that all of the arguments of the correlation function $C[...]$ on the right hand side are now constants (i.e., independent of position $\br$ and time $t$) except for the scaling ratios $|x|(\Lambda r_{_\perp})^{-\zeta}$ and $ |t| (\Lambda r_{_\perp})^{-z}$. therefore, defining
\beq
F(u,w)=\Lambda^{2\chi} C\left[\Lambda^{-1}, u\Lambda^{-\zeta}, w \Lambda^{-z};
m^*, u^*, \Gamma, c_s, c_d, \Lambda \,,
\right]
\label{Fdef}
\eeq
we recover the scaling law equation \rf{crit corr}. 

To obtain the asymptotic forms given in equation \rf{crit corr}, consider first increasing $\rp$ without bound for fixed $x$ and $t$. In this limit, equation \rf{Scaling_Cocriteps}  implies 
\begin{eqnarray}
&&C\left(\brp,x,t; m_0, u_0, \Gamma, c_s, c_d, \Lambda\right)
\nonumber\\
&=&(\Lambda r_{_\perp})^{2\chi}
C\left[\Lambda^{-1}, 0, 0;
m^*, u^*, \Gamma, c_s, c_d, \Lambda
\right]\,.
\nn\\
\label{Scaling_Cocritepsrpinf}
\end{eqnarray}
Since the arguments of  $C\left[\Lambda^{-1}, 0, 0;
m^*, u^*, \Gamma, c_s, c_d, \Lambda
\right]$ are all constants, it should just be a constant itself. This implies $C\left(\brp,x,t; m_0, u_0, \Gamma, c_s, c_d, \Lambda\right)\propto r_{_\perp})^{2\chi}$ in this limit.

For $|x|\to\infty$ with $t$ and $\rp$ fixed, on the other hand, we expect $C\left(\brp,x,t; m_0, u_0, \Gamma, c_s, c_d, \Lambda\right)$ to depend only on $x$. This in turn implies that it must be {\it independent} of $\rp$. From the form of \rf{crit corr}, we see that this can only achieved if, in this limit, $F\left({|x|\over\rp^\zeta}, {|t|\over\rp^z}\right)\propto\left({|x|\over\rp^\zeta}\right)^{2\chi/\zeta}
$, so as to cancel off the $ r_{_\perp}^{2\chi}$ prefactor in \rf{crit corr}. This then implies $C\left(\brp,x,t; m_0, u_0, \Gamma, c_s, c_d, \Lambda\right)
\propto |x|^{2\chi/\zeta}$, as claimed in \rf{crit corr}.

Similar arguments applied to time $t$ imply the large $|t|$ limit in \rf{crit corr}.

\section{Conclusion \& Outlook}{\label{sum}}

In this paper, we have analyzed the universal critical behavior of an incompressible polar active fluid with an easy axis at the order-disorder transition. By using dynamic renormalization group arguments, we map our active, non-equilibrium model onto the equilibrium Ising model with long range dipolar interactions, whose static and dynamic (under model A non-conservative dynamics) behaviors have previously been studied. As a result, we are able to obtain the critical scaling laws for such systems exactly in $d=3$, and  up to $O(\epsilon^2)$ in a $d=3-\epsilon$ expansion in $d=2$.

As in previous studies of incompressible polar active fluid  \cite{chen_natcomm16,chen_pre18}, the success of our mapping onto thermal systems relies on the irrelevance of the advective $\lambda_{1,2}$ terms in the equation of motion \rf{V_x1}. In contrast, in all active systems that have been analyzed so far in which the advective term is relevant in a dry active fluid model, the associated universal behavior has always been distinct from that of any thermal systems \cite{toner_prl95,toner_pre98, toner_prl12, chen_njp15,chen_njp18}. 

Looking ahead, it would be of great interest to identify terms in other non-equilibrium systems that play a similarly key role in revealing the non-equilibrium signature of the systems.

JT thanks the Max Planck Institute for the Physics
of Complex Systems, where the early stage of this work
was performed, for their support. 
LC acknowledges support
by the National Science Foundation of China (under
Grants No.12274452).

%bibliography{C:/Users/chiufanlee/Dropbox/chiufanlee}

\end{document}